\documentclass{siamltex1213}

\usepackage{amsmath}

\usepackage{tikz}

\graphicspath{{./}{Fig/}}

\numberwithin{equation}{subsection}
\newcommand{\B}[6]{[#1_{#2,#3,#4}^{\langle #5 \rangle}]_{#6}}
\newcommand{\GB}[1]{[\chi_{i,k,j}^{\langle y,l \rangle}]_{#1}}
\newcommand{\TP}[4]{\mathcal{P}_{\frac{#1}{#2}}\Bigg(\frac{#3}{#4}\Bigg)}
\newcommand{\snt}[1]{\overrightarrow{#1}} 
\newcommand{\rcv}[1]{\overleftarrow{#1}}
\newcommand{\ovh}[1]{\overline{#1}}
  
\newcommand*\circled[1]{\tikz[baseline=(char.base)]{
            \node[shape=circle,draw,inner sep=0.3pt] (char) {#1};}}
            
\usepackage{color}

 \title{Probabilistic Modeling of IEEE 802.11 Distributed Coordination Functions}

\author{Rui Fang\footnotemark[1] 
\and Zequn Huang\footnotemark[2] 
\and Louis F. Rossi\footnotemark[1]
\and Chien-Chung Shen\footnotemark[2]}

\begin{document}

\maketitle

\renewcommand{\thefootnote}{\fnsymbol{footnote}}

\footnotetext[1]{Department of Mathematical Sciences, University of Delaware}
\footnotetext[2]{Department of Computer and Information Sciences, University of
Delaware}

\begin{abstract}
We introduce and analyze a new Markov model of the IEEE 802.11 Distributed Coordination
Function (DCF) for wireless networks.  The new model is derived from a
detailed DCF description where transition probabilities are
determined by precise estimates of collision probabilities based on network
topology and node states.  For steady state calculations, we approximate joint
probabilities from marginal probabilities using product approximations.  To
assess the quality of the model, we compare detailed equilibrium node states
with results from realistic simulations of wireless networks.  We find very
close correspondence between the model and the simulations in a variety of
representative network topologies.
\end{abstract}

\begin{keywords}
 Wireless networks. Carrier sense multiple access. Hidden
terminal problem. Stochastic modeling. Markov process.
\end{keywords}


\section{Introduction}

Wireless local area networks (WLANs) play a critical role in modern society.
Efficient wireless communications in WLANs require independent nodes to 
coordinate transmissions and receptions of data packets over shared spectrum
so as to mitigate collision.
The coordination of such communications is accomplished through a
Media Access Control (MAC) protocol, a set 
of rules that defines when and how to transmit data from one node to another.
A number of MAC protocols, such as Aloha, CSMA/CD, CSMA/CA, etc.,
have proven to be effective both generally and in special circumstances. 
Most studies of MAC protocols are 
experimental, using either simulated or real network traffic to directly compare 
performance.  MAC protocols themselves are complex and have resisted efforts 
to create consistent mathematical models that can reproduce detailed network 
performance timelines.  The purpose of this paper is to derive and validate a 
predictive mathematical model for the protocol of Carrier Sense Multiple Access
with
Collision Avoidance (CSMA/CA) with binary exponential backoff, which forms the 
IEEE 802.11 Distributed Coordination Function (DCF).
This detailed model is
valuable in and of itself to understand how protocol parameters affect
performance, and it is a natural building block for studying the performance
of upper layer protocols that operate on top of IEEE 802.11 DCF.


All modeling efforts require that we make
assumptions, but the complexity of IEEE 802.11 DCF under general network
topologies requires that investigators make strong assumptions about potential
collisions between nodes.
The seminal work of Bianchi\footnote{Over
6000 citations according to Google Scholar.} \cite{bianchi2000a} on fully
connected single-hop saturated networks begins by
assuming that the collision probability on each node is constant and
independent of network topology and node states.  As we shall see, this is
clearly not the case in general and a Markov model based on this assumption cannot hope to
model DCF.  Numerous works have extended this approach to try to capture
missing elements of DCF in a way that is both
simpler than a full simulation and valuable as a predictive instrument for
studying protocols. 

There have been many extensions of Bianchi's work to model single
hop transmissions where there are no hidden terminals.  
For instance,
the basic model in \cite{bianchi2000a} is adapted to the assumption of
freezing backoff counter due to busy medium in \cite{Ziouva2002}, which is further
polished and strengthened in \cite{Foh2005} by introducing the dependence of
consecutive slots, and also in \cite{Tinnirello2010} by redefining the discrete
time scale given in \cite{bianchi2000a}.
Wu et. al. \cite{Wu2002} augment Bianchi's model by assuming finite retransmission attempts, which is also adopted in \cite{JaehyukChoi2006}.
In \cite{Hadzi-Velkov2003}, the authors propose another model extension for saturation throughput analysis by considering the effect of non-ideal channel conditions, while \cite{Daneshgaran2008} presents a similar model for unsaturated cases.
In addition to throughput analysis, a comprehensive analysis of delay
performance is conducted by \cite{Zhai2003}, where the authors modify node state
transitions in \cite{bianchi2000a} with signal transfer functions to
characterize the probability distribution of MAC layer service time for WLANs in
both saturated and non-saturated traffic situations.
Others, for instance, \cite{Dai2013}, model the statistical behaviors of the
Head-of-Line packet instead of nodes and perform unified study on both
throughput and delay.
A great deal of effort has also been made to model and analyze IEEE 802.11 DCF
in the presence of \textit{hidden terminals}, where some prospective senders are
not within the sensing range of others.
For instance, to model the existence of hidden terminals,
\cite{Wu2006} employs fix-sized time slots and details the state transition
to formalize the channel status considering
the interaction between physical and virtual carrier sensing
in a discrete time Markov system. However,
the authors follow the same assumption that
collision probability is constant regardless of retransmission history. In
contrast, \cite{Jang2012} 
uses the joint backoff stage of the two stations that are hidden from each other
as state in order to account for the interactions between them.  
Unfortunately, these models are
limited to infrastructure scenarios using access points and depend on the network topology.

There has also been some effort to model and analyze multihop
transmissions.  Guillemin et al. propose a 
model for CSMA in multi-hop settings based on a random walk on lattice \cite{Guillemin2011a}.  
The underlying assumption in this model is that node behavior is synchronized so that the problem 
can be parametrized by the queue size on each node.  However, nodes in a
network undergo random exponential backoffs when there is channel contention so
these assumptions are not valid.  Efficiency requires that network
protocols operate asynchronously with each node acting opportunistically to empty its queue or respond to other 
node's requests for it to accept data.  
Other investigators rely upon statistical descriptions of transmission nodes
combing with channel behaviors to develop a model. 
Garetto et al. \cite{Garetto2005} model CSMA for various two contenting flow topologies to study
the unfairness problem and further supplement it to predict throughput in arbitrary topology
\cite{Garetto2008a}. The authors implement a decoupling model for each individual node with an
embedded discrete time renewal process based on the basic assumption that the current channel state
is independent of previous state. However, \cite{Tsertou2008} points out that the above assumption is unrealistic with the presence of hidden terminals and the consequent de-synchronization of the nodes. Instead, Tsertou and Laurenson describe the channel by modeling a first-order dependence between consecutive channel state and adjusted Bianchi's original model using fixed-sized time slot and contention window \cite{Tsertou2008}.
Mustapha et al. \cite{Mustapha2011} apply a discrete-time modeling approach that combines a topology model, a channel model and a simplified node state model with only three states for analyzing throughput of multi-hop ad hoc networks.
In the similar vein but different methodology, Shi et al. \cite{Shi2012} extend Bianchi's
assumptions on backoff-stage dependence of collision probabilities, non-saturate queues, {\em etc.}, and develop a detailed continuous-time model of CSMA networks where the correlations of nodes are described through a companion channel model of joint backoff states. 
Unfortunately, the true statistical description that they are attempting
to capture depends upon network topology and queue sizes.  A more useful model
will generate the statistical description given
network parameters and topology.  This is precisely what we set out to do.

The remainder of the paper is organized as follows. In Section \ref{sec: review} we review the IEEE 802.11 DCF and introduce assumptions used in this paper. In Section \ref{sec: Model} we formulate and discuss the model in details. In Section \ref{sec: example} we apply the model in three basic network configurations and examine the results. Section \ref{sec: conclusion} concludes the paper.

\section{Review of IEEE 802.11 Distributed Coordination Function (DCF)}
\label{sec: review}
In computer networks, a channel access method allows multiple network
devices or nodes to transmit data packets over the same physical
transmission medium ({\em i.e.}, copper wire, air) and share its capacity. The simplest
design is called random access. With this scheme, all network devices may transmit
whenever they want without considering others' conditions.
However, random access leads to packet collisions when two or more devices transmit at the same time.
The resulting mingling of signals will corrupt all data packets involved, and they have to be 
retransmitted at a later time. Hence packet collisions cause lost of information and waste channel bandwidth.

To avoid packet collisions, MAC (Medium Access Control) layer is introduced in the OSI (Open Systems Interconnection) model of computer networks. For Wireless Local Area Networks (WLANs), IEEE 802.11, an international standard, provides a detailed MAC layer specification, in which the fundamental mechanism for network devices to access the medium without any centralized control is called Distributed Coordination Function (DCF).

IEEE 802.11 DCF is a contention based random access scheme, implementing the Carrier Sense Multiple Access with Collision Avoidance (CSMA/CA) protocols. Carrier sense is the ability of a network device to determine if the transmission medium is idle. In general, wireless carrier sense is composed of two distinct techniques: $1)$ CCA (Clear Channel Assessment), which is performed through physical evaluation of the signal energy on the station's radio interface, and $2)$ NAV (Network Allocation Vector), a virtual carrier sense mechanism, which is a data segment that indicates the amount of time required for the transmission immediately following the current packet that contains the NAV.

The collision avoidance feature of CSMA/CA requires that a station transmits only when the channel is sensed to be idle. Unfortunately, collisions may still occur when two stations sense an idle channel at the same instant and subsequently transmit. To reduce the chance of repeated collisions of retransmitted packets, CSMA/CA protocols apply a binary exponential back-off (BEB) algorithm, by which every station selects a random back-off time before each retransmission. The name binary exponential originates from the fact that at each retransmission attempt, the longest possible back-off time doubles. Hence it is less likely for two stations to retransmit at the same moment. 

DCF specifies two approaches for packet transmission. The default scheme is
called Basic Access mechanism. Provided the channel is sensed idle, a sender
transmits the data packet after a random back-off time interval. However, the
data transmission is still vulnerable to packet collision due to the well-known
`hidden terminal problem', or `hidden node problem', in wireless networking. A
node $x$ is called hidden node of node $y$ if $x$ is outside the sensing range
of $y$. A collision may still occur at the receiver node in the presence of
other concurrent transmitters who are hidden from the sender. To address this
issue, DCF provides an optional technique, known as
a Request-to-Send/Clear-to-Send (RTS/CTS) mechanism. Instead of broadcasting a
long and valuable data packet directly, a sender/receiver pair operated in
RTS/CTS mode reserves the channel by handshaking via RTS and CTS short
packets. In particular, since NAV is transmitted along both RTS and CTS packets,
a neighboring node (two nodes are 
neighbors if they can sense each other) overhearing either RTS or CTS packets will defer its own transmission long enough for the addressed communication to finish. Although collisions may still occur between RTS packets, RTS/CTS scheme can reduce the chance of collisions between data packets as long as RTS packets are significantly shorter than the data packets. A more comprehensive description of 802.11 DCF can be found in the standard \cite{IEEE2012}.

\subsection{Preliminaries}
\begin{figure}[!h] 
   \centering
   \includegraphics[width=0.8\textwidth]{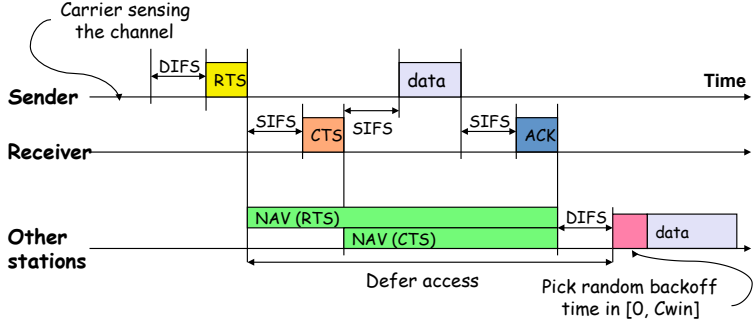} 
   \label{fg: R-C}
   \caption{RTS/CTS Access mechanism}
\end{figure}

In a wireless ad-hoc networks, not all nodes are necessarily within the sensing
range of each other, creating hidden terminals.  To address this, the
802.11 DCF adopts an RTS/CTS/DATA/ACK four way handshaking scheme, shown in Figure \ref{fg: R-C} and described as
follows: 

A sender, $x$, will constantly monitor the channel activity by carrier sensing.
$x$ will not attempt to transmit RTS unless the channel is sensed idle for a
period of time called the Distributed InterFrame Space (DIFS). On the other
hand, $x$ accesses
the channel following the BEB algorithm: at each transmission of RTS packet, the
back-off counter is uniformly chosen between 0 and the current Contention Window
size. Here the contention window determine the longest possible back-off
time a node can choose. The back-off counter is decremented to zero unless $x$
senses a busy channel. This will suspend the counter until the channel is sensed
idle again after a DIFS. Broadcasting of RTS starts when the timer reaches zero.
If the receiver $y$ successfully captures the RTS packet, it will reply to $x$
by
broadcasting a CTS packet after a short period of time interval called the Short
InterFrame Space (SIFS). The contention window will be reset to an initial value
only when $x$ correctly receives the CTS from $y$.  However, CTS reception can
be disrupted by a transmission from another node anywhere within range of $x$.
If the CTS is not received, the contention window doubles, and $x$ retransmits
RTS according to the new contention window after waiting a specified time period
of $T_{out}$, called CTS timeout. Thus, at each failed RTS/CTS handshaking
attempt, $w$ is doubled up to a maximum value. Then the window size remains at
that threshold until it is reset. If the maximum transmission failure limit
(Retry Limit) is reached, $x$ will discard the data packet and the window size
returns to an initial value. The RTS/CTS exchange improves the chances that two
nodes will be able to reserve the channel and exchange data after another SIFS
in a complex environment. At the end of the successful reception of the data
packet from $x$, $y$ immediately responds with a positive acknowledgement (ACK)
after a SIFS. The RTS/CTS/DATA/ACK four way handshaking is complete whenever an
ACK is correctly received by $x$. If not, $x$ will reschedule the data packet
transmission.


\subsection{Assumptions}
To systematically develop a predictive model of 802.11 DCF, we introduce the
following notation and assumptions.
\begin{description}
 \item[Network]: We assume ideal channel conditions.  This means there
will be no noise and the propagation delay is ignored. Each node operates under
homogeneous configurations. All nodes have the same sensing range
$R_s$ and transmission range $R$, where $R < R_s$. 
 
     \item[Timescale]: There exists a constant timescale of least duration, $\sigma$,
\textit{which is equal to the time needed at any node to detect the
transmission of a packet from any other node \cite{bianchi2000a}}. Because $\sigma$ is
very small, we shall assume that any node can immediately detect the
transmission of a packet from any other node inside its sensing range $R_s$. All
the time parameters in the model, i.e, $T_{NAV}$, $T_{busy}$, etc, are assumed
to be multiples $\sigma$.
     
  \item[MAC protocol]: For simplification of modeling, we use a modified version
of IEEE 802.11 DCF implementing the RTS/CTS mechanism: DIFS is set to be one time
unit and SIFS is assumed to be negligible. RTS and CTS packets have the same size,
hence their transmission delays, denoted as $T_{RTS}$ and $T_{CTS}$, are equal.
The protocol still adopt the BEB algorithm and the back-off counter is 
chosen uniformly between 1 and the contention window size. Furthermore, we set
the retry limit of the RTS is the number of times that a contention window
is allowed to double. Hence if the contention window achieves its threshold, we
assume the data packet being sent is dropped. 
  
\item[Data]: There is no retransmission of data packets. A data frame is dropped either because there is a collision at the receiver or retry limit of RTS reached. Also, we assume the acknowledgment packet (ACK) following a successful data packet transmission has fixed size (2 slots) and always succeeds. Hence the transmission time $T_{DATA}$ includes the sending/receiving period of data plus ACK.

\item [Carrier sense]:
\begin{remunerate}
   \item \textbf{CAA - Clear Channel Assessment}:  Since the signals from different neighboring nodes can overlap, the busy period a node physically senses in general will not be constant and will most likely depend on the number of active neighbors. 
     \item \textbf{Network Allocation Vector (NAV)}: It is included in both RTS
and CTS packets indicating how long the channel will be occupied. In the
standard, the value of NAV is $T_{NAVr}=T_{CTS}+T_{DATA}+T_{ACK}$ if contained
in RTS, or $T_{NAVc}=T_{DATA}+T_{ACK}$ if contained in CTS. When a node freezes
through NAV, it will ignore arriving packets until the NAV period ends. On the
other hand, a node will update the freezing period of NAV with the
information overheard from either a CTS or RTS packet if a new NAV value is
greater than the current NAV value. For simplicity, we employ fixed-size NAV
period, and assume a node freezes at the end of NAV if the channel is busy.
   \end{remunerate}

\item [CTS Timeout]: Within the period of CTS timeout, $T_{out}=T_{CTS}+\sigma$, any incoming packets arrived from the physical medium, valid or not, will be ignored. At the end of CTS timeout, we assume a node freezes if the channel is occupied, and resumes back-off/idle if otherwise. 
   \end{description}

\section{Modeling the Distributed Coordination Function}
\label{sec: Model}

In a single hop network (i.e. a fully connected graph), every node can sense each
other and consequently experiences the same level of contention. However, in
a WLAN, the competition among stations for channel access
can be biased: a station with more nodes hidden from it may back off longer or
encounter more packet collisions than the others which have fewer undetectable
contenders. As a result, the performance of the DCF will vary for each node in
the network. 

\subsection{Modeling of node states}
We model each node $x$ in the network as a multi-dimensional stochastic
process, denoted by $$\mathcal{H}_{x}(t):=(s_{x}(t), b_{x}(t), a_{x}(t),
v_{x}(t), \vec{Q}_{x}(t))$$ with the discrete-time Markov chain, in which the
uniform integer time scale, $\sigma$, is adopted: $t_n$ and $t_{n+1}$ correspond
to the beginning of two consecutive slots. ($t_n:=n\sigma$.) 

 \begin{description}
\item [$s_x(t)$]: \textbf{Back-off stage} $(0,1,2, \dots, m)$ of node $x$ at time $t$, where $m$ is the maximum back-off stage. By the exponential back-off scheme described in section \ref{sec: review}, $s_x(t)=i$ implies that the contention window size at time $t  = w_i=2^i w$. $w$ is the initial window size.

\item [$b_x(t)$]: \textbf{Back-off counter} of node $x$ at time $t$. At the beginning of any back-off stage $i$, the counter will randomly choose a value among $(1, \dots, w_i)$ based on the assumptions of protocol. Then for each following time step $t_n$, the back-off counter either decrements or freezes. 

\item [$a_x(t)$]: \textbf{Action/Status} of node $x$ at time $t$:
$$
\left\{
     \begin{array}{lll}
       I, &  x \ \textrm{is idle} &\\
       B, & x\ \textrm{is back-off counting}&\\
       U,&x\ \textrm{is waiting due to unidentified signals sensed}&\\
       R_{\snt{z}},  & x\ \textrm{is sending RTS to } z&\\
       R_{\rcv{z}},  &x\ \textrm{is receiving an uncorrupted RTS from $z$}&\\
       R_{\ovh{z}},  &x\ \textrm{is overhearing an uncorrupted RTS from $z$}&\\
       C_{\snt{z}},  &x\ \textrm{is sending a CTS to } z&\\
       C_{\rcv{z}},  &x\ \textrm{is receiving an uncorrupted CTS from } z&\\
       C_{\ovh{z}},  &x\ \textrm{is overhearing an uncorrupted CTS from $z$}&\\
       A_{\snt{z}},  &x\ \textrm{is sending DATA to } z&\\
       A_{\rcv{z}},  &x\ \textrm{is receiving an uncorrupted DATA from } z&\\
       D_{z},&x\ \textrm{is waiting due to NAV triggered by RTS/CTS from } z&\\ 
       W, &  x\ \textrm{is waiting for a responding CTS}&\\
     \end{array}
   \right.
$$  
Here $z\in N_x$ where $N_x$ denotes the set of neighboring nodes of $x$.
Remark on $W$: $a_x(t)=W$ implies that either the previous RTS packet has been dropped at the receiver so there will be no responding CTS, or the CTS has become unidentified due to collisions at $x$.

 The following table characterizes the actions of $x$ by the behaviors of $x$'s antenna, the channel conditions, and the status of $x$'s queue. For instance, if $a_x(t)=I$, $x$ has nothing to send in the buffer and there is no signal in the medium. Hence its antenna keeps quiet, the channel is sensed free, and its queue is empty. If $a_x(t)=D_z$, $x$ will be frozen because of NAV, which means the antenna is quiet, the channel can be either busy or free depending on the other nodes' actions, and $x$'s queue can be either empty or occupied. The other actions can be described similarly as above.
\begin{table}[htpb]
\footnotesize
  \caption{}\label{t: St-D}
\centering
\begin{tabular} {|  c  || p{2.5cm} | p{2.5cm} | p{2.5cm} |}   
  \hline
  $a_x(t)$ & Antenna (Quiet/Sending) & Channel (Busy/Free) &    Queue \hspace{2cm} (Empty/Occupied)\\ \hline
  $I$ & Quiet & Free &  Empty\\ \hline
  $B$& Quiet & Free & Occupied \\ \hline
      $D_z$ & Quiet&  Busy/Free &  Empty/Occupied\\ \hline
    $W$ &Quiet & Busy/Free & Occupied\\ \hline
   $U/R_{\rcv{z}}/A_{\rcv{z}}/R_{\ovh{z}}/C_{\ovh{z}}$ &Quiet  & Busy & Empty/Occupied \\ \hline
    $C_{\rcv{z}}$ &Quiet & Busy & Occupied\\ \hline
   $C_{\snt{z}}$ & Sending & Busy &  Empty/Occupied\\ \hline
  $R_{\snt{z}}/A_{\snt{z}}$ &Sending &  Busy &  Occupied\\ \hline
\end{tabular}
\end{table}

\item [$v_x(t)$]: \textbf{Virtual timer} associated with $a_x(t)$. It will start
($t = t_0$) at one of the following values and decrement to $0$ at the beginning
of each time slot. Otherwise the timer stays at $0$.
$$
v_x(t_0) = 
\left\{
     \begin{array}{lll}
       t_{RTS}, &  \textrm{if}\ a_x(t)\in\{R_{\snt{z}}, R_{\rcv{z}}, R_{\ovh{z}}\}&\\
       t_{out}, &  \textrm{if}\ a_x(t)= W &\\
       t_{CTS}, & \textrm{if}\ a_x(t)\in\{C_{\snt{z}}, C_{\rcv{z}}, C_{\ovh{z}}\}&\\
       t_{DATA}, &  \textrm{if}\ a_x(t)\in\{A_{\snt{z}}, A_{\rcv{z}}\}&\\
       t_{NAVr}/t_{NAVc}, &  \textrm{if}\  a_x(t)= D_{z} &
     \end{array}
   \right.
$$
where $t_0$ is the initial start time.
Here, $t_{RTS}:=\left \lceil{T_{RTS}/\sigma}\right \rceil -1$ (similarly defined for other time parameters).  

\item [$\vec{Q}_x(t)$]: \textbf{Queue status vector} of node $x$ at time $t$.
Here, $\vec{Q_x}(t)=\langle Y, L \rangle$, where $Y$ is the receiver of the Head of Line (HoL) 
packet that being sent by node $x$. The second entry, $L$, represents the length
of the queue (including the HoL packet) at node $x$. If there
is no packet in the queue, we say $\vec{Q}_x(t)=\vec{0}=\langle \emptyset,0 \rangle$.
Furthermore, we say node $x$ is on $l$-th layer at time $t$ if
$L=l$. Whenever the node $x$ successfully receives a
packet during the back-off counting, $L$ is increased by $1$. If node $x$
finishes transmitting a packet (either success or failure), $L$ is dropped by
$1$, and $Y$ will be updated based on the receiver of the next packet in the
queue. 
\end{description}

\subsection{Modeling of States Transitions}
\label{sec: transit}
\subsubsection{x as a listener/receiver}
\begin{figure}[h] 
   \centering
   \includegraphics[width=\textwidth]{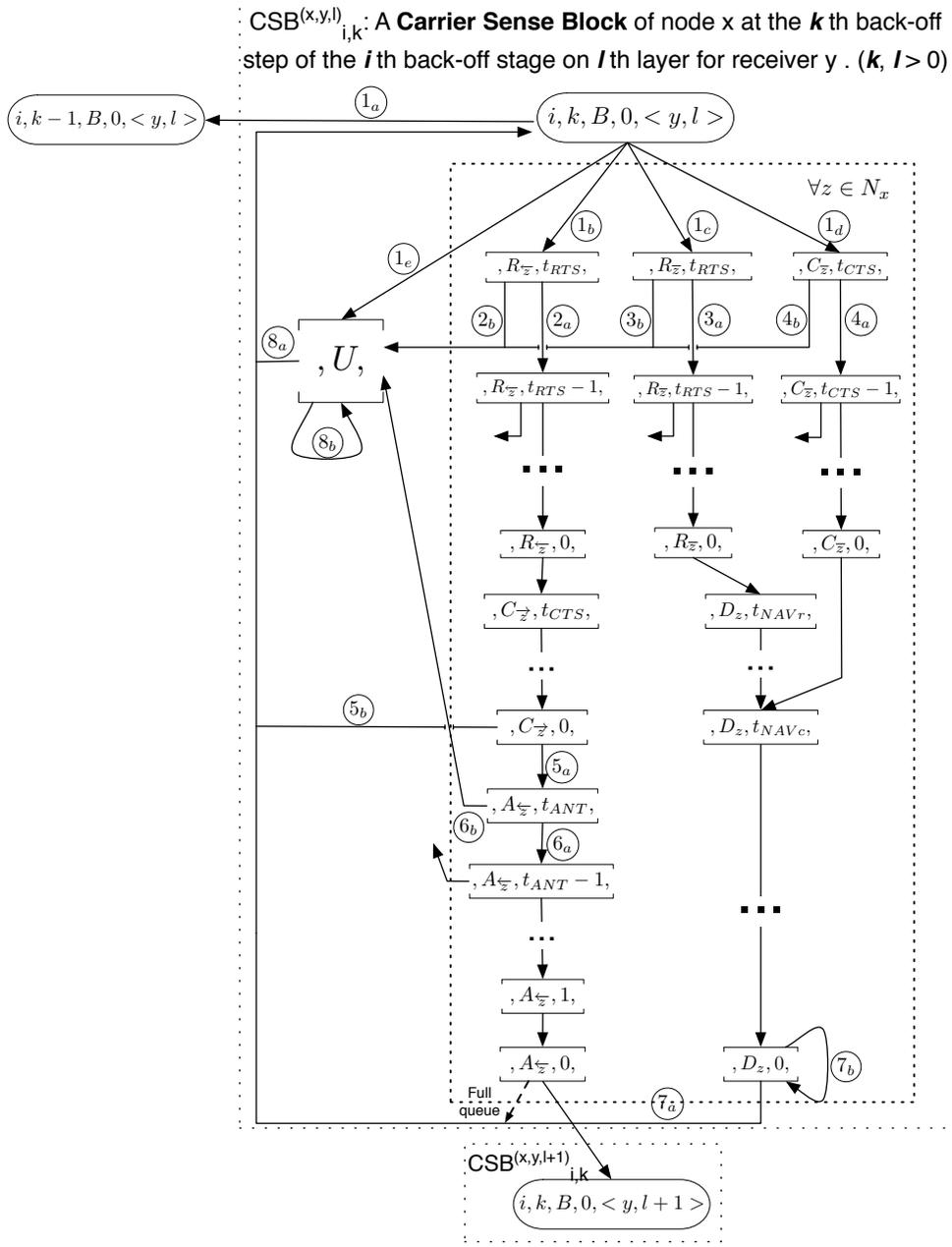} 
   \caption{Carrier Sense Block at upper layer}
   \label{csb1}
\end{figure}

\begin{figure}[h] 
   \centering
   \includegraphics[width=0.75\textwidth]{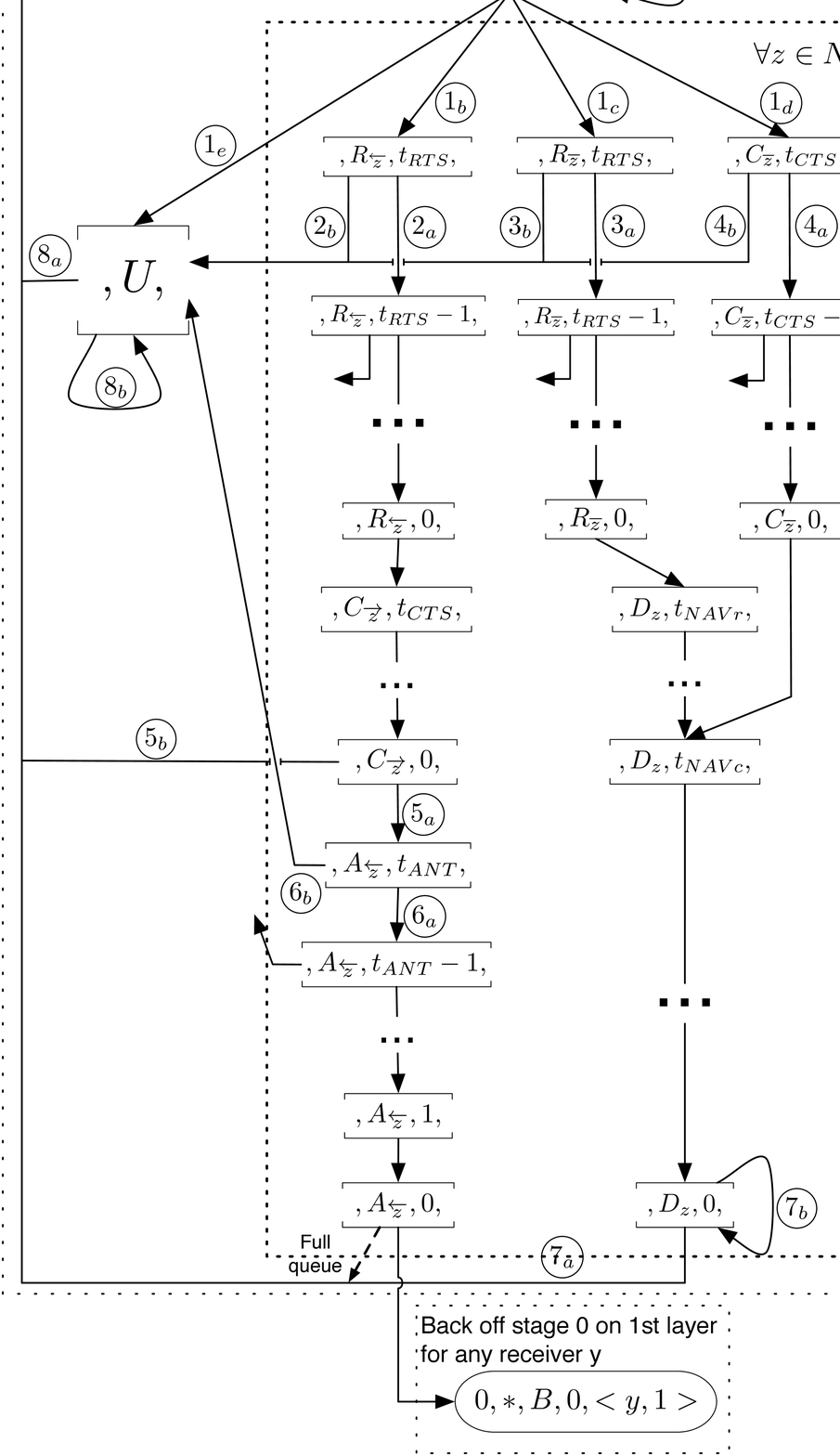} 
   \caption{Carrier Sense Block at base layer}
   \label{csb2}
\end{figure}

A node $x$ is a listener when it is in back-off counting (with occupied queue)
or idle (with empty queue). It consistently monitors the channel by both
physical and virtual carrier sense. Upon the successful reception of a RTS
packet, $x$ becomes a receiver by completing the RTS/CTS/DATA/ACK handshake.
Diagram \ref{csb1} and \ref{csb2} represent the states' transitions for $x$
based on the description of 802.11 DCF and the assumptions in section \ref{sec: review}. Both
diagrams share a similar structure, called \textbf{Carrier Sense Block} (CSB),
which repeatedly appears in our model for every pair of back-off stage and
back-off counter. 

For Figure \ref{csb1}, suppose that at time step $t_n$ where $n=0,1,2,\cdots$,
node $x$ is at the $k$th step of the $i$th backoff stage for receiver $y$ with
$l$ packets in the queue. At the next time step there are five possible state
transitions on node $x$, associated with the following probabilities
respectively: 
\begin{eqnarray}
\circled{$1_a$}&=&\textrm{Prob}\big\{(i,k-1,B,0,\langle y,l \rangle)_{n+1}\big|(i,k,B,0,\langle y,l \rangle)_{n} \big\}\\
\circled{$1_b$}&=&\textrm{Prob}\big\{(i,k,R_{\rcv{z}},t_{RTS},\langle y,l \rangle)_{n+1}\big|(i,k,B,0,\langle y,l \rangle)_{n} \big\}\\
\circled{$1_c$}&=&\textrm{Prob}\big\{(i,k,R_{\ovh{z}},t_{RTS},\langle y,l \rangle)_{n+1}\big|(i,k,B,0,\langle y,l \rangle)_{n} \big\}\\
\circled{$1_d$}&=&\textrm{Prob}\big\{(i,k,C_{\ovh{z}},t_{CTS},\langle y,l \rangle)_{n+1}\big|(i,k,B,0,\langle y,l \rangle)_{n} \big\}\\
\circled{$1_e$}&=&\textrm{Prob}\big\{(i,k,U,0,\langle y,l \rangle)_{n+1}\big|(i,k,B,0,\langle y,l \rangle)_{n} \big\}
\end{eqnarray}
Here we adopt the short notion: 
\begin{eqnarray*}
&&P\{(z_1, z_2, z_3, z_4, z_5)_{n+1}\big|(z'_1, z'_2, z'_3, z'_4, z'_5)_{n}\}\\ &=& P\{\mathcal{H}_x(t_{n+1})=(z_1, z_2, z_3, z_4, z_5)\big|\mathcal{H}_x(t_n)=(z'_1, z'_2, z'_3, z'_4, z'_5)\}
\end{eqnarray*}
Transition \circled{$1_a$} occurs when $x$ detects a quiet channel, that is,
currently no neighbors of $x$ are broadcasting or beginning to transmit any
signals. As a result, the back off counter decrements by $1$. Transition
\circled{$1_b$} accounts for the fact that, one of $x$'s neighbor, $z$, begins
to send a RTS packet for $x$ while others neighboring nodes stay quiet. In this
case, node $x$ takes the first step of receiving the RTS packet, so that
$a_x(t_{n+1})=R_{\rcv{z}}, v_{x}(t_{n+1})=t_{RTS}$. 
Transition \circled{$1_c$} or \circled{$1_d$} takes place provided that only $z$ starts to broadcast a RTS packet or a CTS packet not for $x$. In those scenarios, $a_x(t_{n+1})=R_{\ovh{z}}, v_{x}(t_{n+1})=t_{RTS}$ or $a_x(t_{n+1})=C_{\ovh{z}}, v_{x}(t_{n+1})=t_{CTS}$. The transition \circled{$1_e$}, $a_x(t_{n+1})=U$, happens when $x$ detects disordered signals in the channel, caused by either corrupted or partial packets from $x$'s neighbors. 
 
During the receiving (overhearing) of RTS or CTS from a neighbor $z$, node $x$ may observe packet collisions when the hidden nodes of $z$ initiate transmissions to $x$. Thus, given the $j$-th step of receiving ($v_x(t_n)=j$), we have the following probabilities associated with the transitions \circled{$2_b$}, \circled{$3_b$} and \circled{$4_b$}:
\begin{eqnarray}
\circled{$2_b$}&=&\textrm{Prob}\big\{\big (i,k,U,0,\langle y,l \rangle)_{n+1}|(i,k,R_{\rcv{z}},j,\langle y,l \rangle)_{n} \big\}\\
\circled{$3_b$}&=&\textrm{Prob}\big\{(i,k,U,0,\langle y,l \rangle)_{n+1}\big|(i,k,R_{\ovh{z}},j,\langle y,l \rangle)_{n}\big\}\\
\circled{$4_b$}&=&\textrm{Prob}\big\{(i,k,U,0,\langle y,l \rangle)_{n+1}\big|(i,k,C_{\ovh{z}},j,\langle y,l \rangle)_{n}\big\}
\end{eqnarray}
Otherwise, $x$ keeps receiving and the virtual counter $v_x(t)$ decreases by $1$ at each time step with the probabilities:
\begin{eqnarray}
\circled{$2_a$}&=&\textrm{Prob}\big\{\big (i,k,R_{\rcv{z}},j-1,0,\langle y,l \rangle)_{n+1}|(i,k,R_{\rcv{z}},j,\langle y,l \rangle)_{n} \big\}\\
\circled{$3_a$}&=&\textrm{Prob}\big\{(i,k,R_{\ovh{z}},j-1,0,\langle y,l \rangle)_{n+1}\big|(i,k,R_{\ovh{z}},j,\langle y,l \rangle)_{n}\big\}\\
\circled{$4_a$}&=&\textrm{Prob}\big\{(i,k,C_{\ovh{z}},j-1,0,\langle y,l \rangle)_{n+1}\big|(i,k,C_{\ovh{z}},j,\langle y,l \rangle)_{n}\big\}
\end{eqnarray}

If a RTS is successfully received, that is, $a_x(t_n)=R_{\rcv{z}}, v_x(t_n)=0$,
$x$ will start to respond with a CTS to $z$, shown by
$a_x(t_{n+1})=C_{\snt{z}}$, $v_{x}(t_{n+1})=t_{CTS}$. The transmission of the
CTS takes $t_{CTS}$ steps and if successful, $x$ should begin to receive a data
packet from $z$. Otherwise, no data will be sent, 
and $x$ resumes carrier sensing. Thus we have the following transition
probabilities:
\begin{eqnarray}
\circled{$5_a$} &=&\textrm{Prob}\big\{\big (i,k, A_{\rcv{z}}, t_{DATA},\langle y,l \rangle)_{n+1}|(i,k, C_{\snt{z}},0,\langle y,l \rangle)_{n} \big\}\\
\circled{$5_b$}&=&\textrm{Prob}\big\{\big (i,k, B, 0,\langle y,l \rangle)_{n+1}|(i,k, C_{\snt{z}},0,\langle y,l \rangle)_{n} \big\}
\end{eqnarray}
At each step of receiving DATA, there are two possible transitions:
\begin{eqnarray}
\circled{$6_a$}&=&\textrm{Prob}\big\{\big (i,k, A_{\rcv{z}}, j-1,\langle y,l \rangle)_{n+1}|(i,k, A_{\rcv{z}},j,\langle y,l \rangle)_{n} \big\}\\
\circled{$6_b$}&=&\textrm{Prob}\big\{\big (i,k, U, 0,\langle y,l \rangle)_{n+1}|(i,k, A_{\rcv{z}},j,\langle y,l \rangle)_{n} \big\}
\end{eqnarray}
For the first transition, $x$ correctly receives the next piece of data so
$v_x(t)$ decrease by 1. Otherwise, $x$ detects a collision, which implies the
signal is corrupted, shown by $a_x(t_{n+1})=U$. When $v_{x}(t)=1$, the receiving
of data is complete and $x$ shall reply with an ACK packet. When $v_x(t)$
decreases to $0$, that is, the DATA/ACK handshake is successful, $x$ will resume
back off counting on the next layer and the queue size increases by 1. If the
queue is full, as shown by the dashed arrow in diagram \ref{csb1}, the data
received will be dropped and $x$ will resume back-off counting on the same
layer. 

If $x$ successfully overhears a RTS, then with probability $1$ it will go to
silent mode $D_{z}$ and update $v_x(t)$ to $t_{NAVr}$. Similarly, if a CTS
is overheard, $v_{x}(t)$ changes to $t_{NAVc}$. Upon $v_x(t)$ reaches $0$, the
behavior of $x$ at the next time step depends on the channel status. With
probability \circled{$7_a$}, $x$ resumes back-off counting because it senses a
quiet channel, or with probability \circled{$7_b$}, $x$ detects a busy channel
and waits.
\begin{eqnarray}
\circled{$7_a$}&=&\textrm{Prob}\big\{\big (i,k,B, 0,\langle y,l \rangle)_{n+1}|(i,k,D_z,0,\langle y,l \rangle)_{n} \big\}\\
\circled{$7_b$}&=&\textrm{Prob}\big\{\big (i,k,D_z,0,\langle y,l \rangle)_{n+1}|(i,k,D_z,0,\langle y,l \rangle)_{n} \big\}
\end{eqnarray}

Finally, if $x$ senses jumbled signals in the channel at time step $t_n$
($a_x(t_n)=U$), then after one discrete time step $x$ either senses the channel
is clear and resumes back-off counting ($a_x(t_{n+1})=B$), or detects a busy
channel ($a_x(t_{n+1})=U$) and waits, with the following probabilities: 
\begin{eqnarray}
\circled{$8_a$}&=&\textrm{Prob}\big\{\big (i,k,B, 0,\langle y,l \rangle)_{n+1}|(i,k,U,0,\langle y,l \rangle)_{n} \big\}\\
\circled{$8_b$}&=&\textrm{Prob}\big\{\big (i,k,U, 0,\langle y,l \rangle)_{n+1}|(i,k,U,0,\langle y,l \rangle)_{n} \big\}
\end{eqnarray}

For Figure \ref{csb2} where $x$ has empty queue, the state transitions are
similar except with probability \circled{$1_a$} x stays idle and keeps
monitoring the channel. After a data packet is received, if $x$ is a relay node,
it will randomly or deterministically choose a receiver in $N_x$ and set a
back-off counter between $1$ and the initial contention window size $w$.

\subsubsection{$x$ as a sender}
\begin{figure}[h] 
   \centering
   \includegraphics[width=0.9\textwidth]{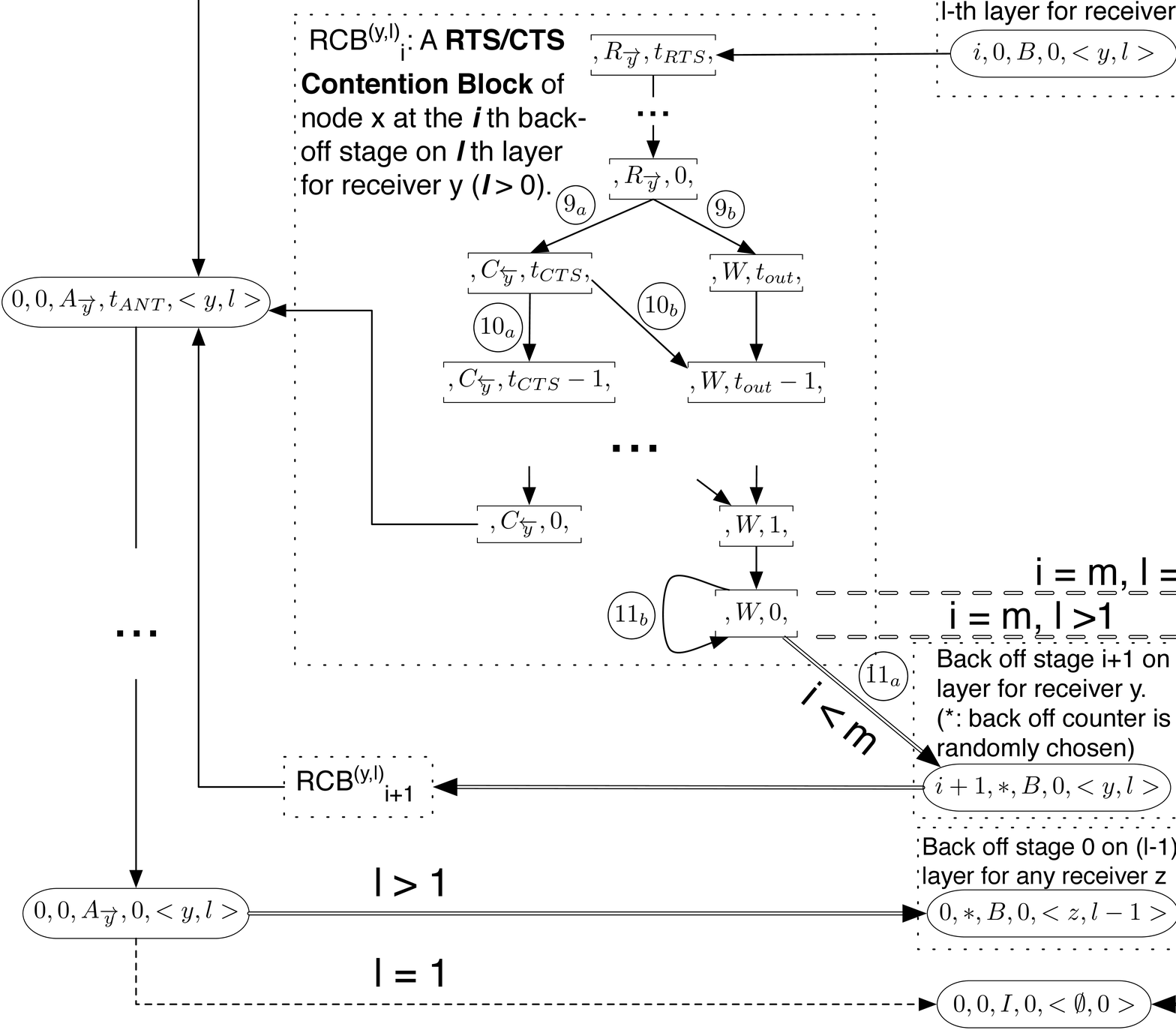} 
   \caption{RTS/CTS Contention Block}
   \label{rcb}
\end{figure}

At the end of counting ($b_x(t)=0$) at any back-off stage, $x$ becomes a sender
by immediately initiating a RTS transmission. The state transitions of $x$ as a
sender are shown in Figure \ref{rcb}. A structure, called a \textbf{RTS/CTS
Contention Block} (RCB) emerges in the model whenever $x$ attempts a RTS/CTS
handshake. 

Suppose node $x$ transits a RTS packet to y during $i$th backoff stage with $l$ packets in the queue. After a time period of $t_{RTS}$,  the RTS transmission either succeeds and begins to receive a CTS from $y$ with probability
\begin{equation}
\circled{$9_a$}=\textrm{Prob}\big\{\big (i,0,C_{\rcv{y}}, t_{CTS},\langle y,l \rangle)_{n+1}|(i,0,R_{\snt{y}},0,\langle y,l \rangle)_{n} \big\}
\end{equation}
or fails with probability
\begin{equation}
\circled{$9_b$} = \textrm{Prob}\big\{\big (i,0,W,t_{out},\langle y,l \rangle)_{n+1}|(i,0,R_{\snt{y}},0,\langle y,l \rangle)_{n} \big\}
\end{equation}
In this case, there will be no reply so that $x$ waits until the virtual counter $v_x(t)$ reaches $0$. 

At each step of receiving a CTS, depending on whether there is a collision at
$x$, we have the following transition probabilities:
\begin{eqnarray*}
{\small\circled{$10_a$}}&=&\textrm{Prob}\big\{\big (i,0,C_{\rcv{y}}, j-1,\langle y,l \rangle)_{n+1}|(i,0,C_{\rcv{y}}, j ,\langle y,l \rangle)_{n} \big\}\\
{\small\circled{$10_b$}}&=&\textrm{Prob}\big\{\big (i,0,W, t_{out}-j,\langle y,l \rangle)_{n+1}|(i,0,C_{\rcv{y}}, j ,\langle y,l \rangle)_{n} \big\}
\end{eqnarray*}

When the receiving of the CTS is complete, $x$ will initiate an DATA/ACK
handshake, which lasts $t_{DATA}$ time steps. In the end, if $l=1$, i.e. the
queue is empty, $x$ becomes idle, otherwise $x$ restarts the back-off
procedure for the next HoL packet and the queue size decreases by $1$.

Finally, suppose the RTS/CTS handshake fails, $x$ senses the channel at the end
of the CTS timeout. Given no transmitting neighbors, if the current back-off
stage is less than the maximum stage allowed ($i<m$), $x$ will reset the back
off counter between $1$ and the doubled contention window size, then resume
counting procedure at the back off stage $i+1$. The associated probability
function is:
\begin{eqnarray*}
{\small\circled{$11_a$}}&=& \sum_{k}\textrm{Prob}\big\{\big (i+1,k,B, 0,\langle y,l \rangle)_{n+1}|(i,0,W, 0 ,\langle y,l \rangle)_{n} \big\}
\end{eqnarray*}
However, if the maximum stage is reached, then the data packet will be dropped. Based on the current queue size, $x$ can either restart back-off procedure ($l>1$) or become idle ($l=1$):
\begin{eqnarray*}
{\small\circled{$11_a$}}&=& \sum_{k}\textrm{Prob}\big\{\big (0,k,B, 0,\langle y,l-1\rangle)_{n+1}|(m,0,W, 0 ,\langle y,l \rangle)_{n} \big\}
\end{eqnarray*}
\begin{eqnarray*}
{\small\circled{$11_a$}}&=&\textrm{Prob}\big\{\big (0,0,I, 0,\langle \emptyset, 0\rangle)_{n+1}|(m,0,W, 0 ,\langle y,1\rangle)_{n} \big\}
\end{eqnarray*}
For the case that a busy channel is sensed, $x$ will freeze, as shown by,
\begin{eqnarray*}
{\small\circled{$11_b$}}&=&\textrm{Prob}\big\{\big (i,0,W, 0,\langle y,l \rangle)_{n+1}|(i,0,W, 0 ,\langle y,l \rangle)_{n} \big\}
\end{eqnarray*}


\subsection{Representation of Transition Probabilities}
\label{subsec: represent}
In this section, we address the formulations of transition probability functions in detail.
For simplicity, we first denote the probability density function for any node $x$ in the network at time step $t_n$ by
$$ P^{(n)}([\chi_{i,k,j}^{\langle y,l \rangle}]_x) :=\text{Prob}\{\mathcal{H}_{x}(t_n)=\mathbf{h}_x\}=\text{Prob}\{\mathcal{H}_{x}(t_n)=(i,k,\chi, j, \langle y,l\rangle)\}$$
Here $i\in [0,m]$, $k\in [0, 2^iw]$, $\chi \in \{I, B, U, R_{\snt{z}/\rcv{z}/\ovh{z}}, C_{\snt{z}/\rcv{z}/\ovh{z}}, A_{\snt{z}/\rcv{z}}, D_z, W\}$, $j\in [0, t_{NAVr}]$, $y,z\in N_x$ and $l\in [0, L_x]$ where $N_x$ is the set that contains $x$'s neighbors (and $\emptyset$), and $L_x$ represents the maximum queue size of $x$. 
The joint probability density functions are similarly defined and symmetric:
\begin{equation*}
P^{(n)}([\chi_{i,k,j}^{\langle y,l \rangle}]_{x}, [\bar{\chi}_{i',k',j'}^{\langle y',l' \rangle}]_{x'}, \cdots )=P^{(n)}([\chi_{i,k,j}^{\langle y,l \rangle}]_{x'}, [\bar{\chi}_{i',k',j'}^{\langle y',l' \rangle}]_{x}, \cdots )
\end{equation*}

The probability density function of node $x$ can be obtained by marginalizing out other nodes in the joint state probability density function, i.e.
\begin{align*}
\textrm{Prob}(\mathcal{H}_{x}(t_n)=\mathbf{h}_x)&=\textrm{Prob}(\mathcal{H}_{x}(t_n)=\mathbf{h}_{x}, \bullet )\\
&=\sum_{(\mathbf{h}_x',\cdots)\in \Omega(\mathbf{h}_x;x',\cdots)}\textrm{Prob}(\mathcal{H}_{x}(t_n)=\mathbf{h}_x, \mathcal{H}_{x}(t_n)=\mathbf{h}_{x'},\cdots) 
\end{align*}
where $\Omega(\mathbf{h}_x;x',\cdots)$ represents the sub state space of nodes \{$x',\cdots$\} such that, 
$$
\textrm{Prob}(\mathcal{H}_{x}(t_n)=\mathbf{h}_x, \mathcal{H}_{x'}(t_n)=\mathbf{h}_{x'},\cdots) \not \equiv 0
$$
$\forall (\mathbf{h_{x'}},\cdots)\in \Omega(\mathbf{h}_x;x',\cdots)$

On the other hand, given a set of marginal densities, the joint distribution in
general cannot be uniquely determined unless the random variables are
independent. This brings forward the main challenge in our modeling framework
since for each node, all the critical state transitions mentioned in the last
section are dependent on the concurrent states of its neighboring nodes. To be
precise, suppose $N_x=\{x_1, x_2, \cdots, x_r\}$ and expanding the marginal
probability density function of $x$ on $\Omega(\mathbf{h}_x;x_1,\cdots, x_r)$,
we have
\begin{align*}
&\qquad \textrm{Prob}\{\mathcal{H}_x(t_{n+1})=\mathbf{h}'_x|\mathcal{H}_x(t_n)=\mathbf{h}_x\}\\
&=\sum_{(\mathbf{h}_{x_1}\cdots,\mathbf{h_{x_r}})\in \Omega_{\mathcal{A}(\mathbf{h}_x; x_1,\cdots,x_r)}}\frac{\textrm{Prob}\{\mathcal{H}_x(t_n)=\mathbf{h}_x, \mathcal{H}_{x_1}(t_n)=\mathbf{h}_{x_1},\cdots,\mathcal{H}_{x_r}(t_n)=\mathbf{h}_{x_r}\}}{\textrm{Prob}\{\mathcal{H}_x(t_n)=\mathbf{h}_x\}}\\
&=\frac{\sum_{\Omega_{\mathcal{A}}(\B{\chi}{i}{k}{j}{y,l}{x}; x_1, \cdots,
x_r)}P^{(n)}(\B{\chi}{i}{k}{j}{y,l}{x}, \B{\chi}{i}{k}{j}{y,l}{x_1}, \cdots,
\B{\chi}{i}{k}{j}{y,l}{x_r})}{\sum_{\Omega(\B{\chi}{i}{k}{j}{y,l}{x}; x_1,
\cdots, x_r)}P^{(n)}(\B{\chi}{i}{k}{j}{y,l}{x}, \B{\chi}{i}{k}{j}{y,l}{x_1},
\cdots, \B{\chi}{i}{k}{j}{y,l}{x_r})}
:=
\frac{\mathcal{F}_{\Omega_{\mathcal{A}}}(\B{\chi}{i}{k}{j}{y,l}{x})}{\mathcal{
F}_{\Omega}(\B{\chi}{i}{k}{j}{y,l}{x})}
\end{align*}
where $\Omega_{\mathcal{A}}(\mathbf{h}_x;x_1,\cdots, x_r)\subseteq \Omega(\mathbf{h}_x;x_1,\cdots, x_r)$ and 
\begin{eqnarray*}
&&\textrm{Prob}\{\mathcal{H}_x(t_{n+1})=\mathbf{h}'_x|\mathcal{H}_x(t_n)=\mathbf{h}_x, \mathcal{H}_{x_1}(t_n)=\mathbf{h}_{x_1},\cdots,\mathcal{H}_{x_r}(t_n)=\mathbf{h}_{x_r}\}\\
&=&
\begin{cases}
    1,  & \textrm{if } (\mathbf{h}_{x_1}\cdots,\mathbf{h_{x_r}})\in \Omega_{\mathcal{A}}(\mathbf{h}_x; x_1,\cdots,x_r)\\
    0,  & \text{otherwise}
\end{cases}
\end{eqnarray*}
For the purpose of evaluating the transition probability functions introduced in
Section \ref{sec: transit}, we shall establish their connections (shown by functions
$\mathcal{F}_{\Omega}$ and $\mathcal{F}_{\Omega_{\mathcal{A}}}$) to the
probability density functions of joint states with the neighboring nodes. The
joint state spaces $\Omega$ and $\Omega_{\mathcal{A}}$ will be discussed based
on four categories of actions $\mathcal{A}$ that $x$ takes.

\subsubsection{Carrier sensing while in the idle or back-off states} 
Let us suppose at the current time step $t_n$ $x$ is sensing a free channel and
not freezing or waiting, that is, $x$ is in back off state $B$ (or equivalently,
idle state $I$, if its queue is empty), and the parameters $i',k',y',l'$ are
fixed: $\mathcal{H}_x(t_n)=(i',k',B,0,\langle y',l' \rangle)$. Referring to
Table \ref{t: St-D} the channel must be quiet, hence all the neighboring nodes
of $x$ are not sending and not receiving from $x$ or common neighbors with
$x$ (as $x$ is known to be in the back-off state). Using the notations of
cartesian product, we then have
\begin{align*}
&\qquad\Omega(\B{B}{i'}{k'}{0}{y',l'}{x}; x_1,\cdots, x_r)=\Omega(\B{B}{i'}{k'}{0}{y',l'}{x}; x_1)\times\cdots \times \Omega(\B{B}{i'}{k'}{0}{y',l'}{x}; x_r)\\
 &=\times_{x_{\alpha}\in N_x}\Omega(\B{B}{i'}{k'}{0}{y',l'}{x}; x_\alpha)\\
 &=\times_{x_{\alpha}\in N_x}\{\mathcal{H}_{x_\alpha}(t_n)|\underbrace{\chi_\alpha \notin \{R_{\snt{z}}, C_{\snt{z}}, A_{\snt{z}}\}}_{\textit{not transmitting}} \& \underbrace{\chi_{\alpha} \notin \{R_{\rcv{z'}/\ovh{z'}}, C_{\rcv{z'}/\ovh{z'}}, A_{\rcv{z'}}, D_{z'}\}, z'\notin N_x}_{\textit{not interacting with $x$ and $N_x$}}\}
\end{align*}
such that $P^{(n)}(\B{B}{i'}{k'}{0}{y',l'}{x}))=\mathcal{F}_{\Omega}(\B{B}{i'}{k'}{0}{y',l'}{x}))$.
At the next time step $t_{n+1}$, if no neighbors of $x$ are ready to send any signals, the channel will remain quiet. Hence we conclude that
\begin{eqnarray*}
\circled{$1_a$}=\frac{\mathcal{F}_{\Omega_{1a}}(\B{B}{i'}{k'}{0}{y',l'}{x})}{\mathcal{F}_{\Omega}(\B{B}{i'}{k'}{0}{y',l'}{x})}=\frac{\sum_{\Omega_{1a}(\B{B}{i'}{k'}{0}{y',l'}{x}; x_1, \cdots, x_r)}P^{(n)}([B_{i',k',0}^{\langle y',l' \rangle}]_x, [\chi_{i,k,j}^{\langle y,l \rangle}]_{x_1}, \cdots, [\chi_{i,k,j}^{\langle y,l \rangle}]_{x_r})}{\sum_{\Omega(\B{B}{i'}{k'}{0}{y',l'}{x}; x_1, \cdots, x_r)}P^{(n)}([B_{i',k',0}^{\langle y',l' \rangle}]_x,[\chi_{i,k,j}^{\langle y,l \rangle}]_{x_1}, \cdots, [\chi_{i,k,j}^{\langle y,l \rangle}]_{x_r})}
\end{eqnarray*}
Here, $\Omega_{1a}(\B{B}{i'}{k'}{0}{y',l'}{x}; x_1, \cdots, x_r)\subseteq \Omega(\B{B}{i'}{k'}{0}{y',l'}{x}; x_1, \cdots, x_r)$ and includes an extra restriction:
\begin{align*}
&\qquad \Omega_{1a}(\B{B}{i'}{k'}{0}{y',l'}{x}; x_1, \cdots, x_r) = \times_{x_\alpha\in N_x}\Omega_{1a}(\B{B}{i'}{k'}{0}{y',l'}{x}; x_\alpha)\\
&= \times_{x_{\alpha}\in N_x}\{\mathcal{H}_{x_\alpha}(t_n)\in
\Omega(\B{B}{i'}{k'}{0}{y',l'}{x}; x_\alpha)|\underbrace{(\chi, k, j)_{x_\alpha}
\notin\{(B, 0, 0),(R_{\rcv{z}},k, 0),(C_{\rcv{z}}, 0, 0)\}}_{\textit{not begin
to send RTS/CTS/DATA}}\}.
\end{align*}
For transition \circled{$1_b$}, it accounts for the fact that one neighbor of $x$, for example, $x'$, begins to send a RTS packet to $x$, while the rest neighbors do not begin to send. We thus have
\begin{align*}
&\qquad \Omega_{1b}(\B{B}{i'}{k'}{0}{y',l'}{x}; x', \cdots, x_r) = \Omega_{1b}(\B{B}{i'}{k'}{0}{y',l'}{x}; x')\times_{x_\alpha\in N_x\backslash x'}\Omega_{1a}(\B{B}{i'}{k'}{0}{y',l'}{x}; x_\alpha)\\
&= \{\mathcal{H}_{x'}(t_n)|\underbrace{(\chi, k, y)_{x'}=(B, 0, x)}_{\textit{begins to sent a RTS to x}}\}\times_{x_\alpha\in N_x\backslash x'}\Omega_{1a}(\B{B}{i'}{k'}{0}{y',l'}{x}; x_\alpha)
\end{align*}
such that
\begin{eqnarray*}
\circled{$1_b$}=\frac{\mathcal{F}_{\Omega_{1b},x'}(\B{B}{i'}{k'}{0}{y',l'}{x})}{
\mathcal{F}_{\Omega}(\B{B}{i'}{k'}{0}{y',l'}{x})}.
\end{eqnarray*}
On the other hand, if $x'$ begins to send a RTS not to $x$ while all other
neighbors remain quiet and do not start to transmit any packet, $x$ will start
to overhear the  RTS. The probability \circled{$1_c$} is given by
\begin{eqnarray*}
\circled{$1_c$}=\frac{\mathcal{F}_{\Omega_{1c},x'}(\B{B}{i'}{k'}{0}{y',l'}{x},x')}{\mathcal{F}_{\Omega}(\B{B}{i'}{k'}{0}{y',l'}{x})}
\end{eqnarray*}
where $\Omega_{1c}(\B{B}{i'}{k'}{0}{y',l'}{x}; x', \cdots, x_{r})$ is similarly defined by
\begin{align*}
&\qquad \Omega_{1c}(\B{B}{i'}{k'}{0}{y',l'}{x}; x', \cdots, x_r) \\
&= \{\mathcal{H}_{x'}(t_n)|\underbrace{(\chi, k)_{x'}=(B, 0)\ \& \ y_{x'}\ne x}_{\textit{begins to sent a RTS not to x}}\}\times_{x_\alpha\in N_x\backslash x'}\Omega_{1a}(\B{B}{i'}{k'}{0}{y',l'}{x}; x_\alpha)
\end{align*}

Likewise, if $x'$ starts to sent a CTS not to $x$ while the remaining neighbors stay quiet and do not initiate a transmission, we get
\begin{align*}
&\qquad \Omega_{1d}(\B{B}{i'}{k'}{0}{y',l'}{x}; x', \cdots, x_r) \\
&= \{\mathcal{H}_{x'}(t_n)|\underbrace{(\chi, k)_{x'}=(R_{\rcv{z}}, 0), z\notin N_x}_{\textit{begins to sent a CTS not to $x$ (or $N_x$)}}\}\times_{x_\alpha\in N_x\backslash x'}\Omega_{1a}(\B{B}{i'}{k'}{0}{y',l'}{x}; x_\alpha)
\end{align*}
so that 
\begin{eqnarray*}
\circled{$1_d$}=\frac{\mathcal{F}_{\Omega_{1d},x'}(\B{B}{i'}{k'}{0}{y',l'}{x})}{\mathcal{F}_{\Omega}(\B{B}{i'}{k'}{0}{y',l'}{x})}
\end{eqnarray*}

Otherwise, $x$ detects an unidentified busy channel. The corresponding transition has probability computed by 
\begin{multline*}
\circled{$1_e$}=1-\frac{\mathcal{F}_{\Omega_{1a}}(\B{B}{i'}{k'}{0}{y',l'}{x})}{\mathcal{F}_{\Omega}(\B{B}{i'}{k'}{0}{y',l'}{x})}-\sum_{x'\in N_x}\Bigg(\frac{\mathcal{F}_{\Omega_{1b},x'}(\B{B}{i'}{k'}{0}{y',l'}{x})}{\mathcal{F}_{\Omega}(\B{B}{i'}{k'}{0}{y',l'}{x})}+\frac{\mathcal{F}_{\Omega_{1c},x'}(\B{B}{i'}{k'}{0}{y',l'}{x})}{\mathcal{F}_{\Omega}(\B{B}{i'}{k'}{0}{y',l'}{x})}\\
+\frac{\mathcal{F}_{\Omega_{1d},x'}(\B{B}{i'}{k'}{0}{y',l'}{x})}{\mathcal{F}_{\Omega}(\B{B}{i'}{k'}{0}{y',l'}{x})}\Bigg)
\end{multline*}

\subsubsection{Receiving/overhearing packets}
Next suppose at $t_n$ $x$ is receiving or overhearing a packet from a neighbor $x'$ without interference by the others that are hidden from $x'$: 
$\mathcal{H}_x(t_n)=(i',k',\tilde{\chi},j',\langle y',l' \rangle)$,
$\tilde{\chi}\in \{R_{\rcv{x'}},
R_{\ovh{x'}},C_{\rcv{x'}},C_{\ovh{x'}},A_{\rcv{x'}}\}$, $j'\ne 0$. We observe
that $x'$ is at the $j'$-th step of transmitting the same packet, and all the
other neighbors of $x$ that are hidden from $x'$ are quiet and do not interact
with $x$. The common neighbors of $x$ and $x'$ are ignored because they share
the same channel and will not intervene. Now assume
$N_{xx'}:=\{x_1,x_2,\cdots,x_h\}$ are hidden from $x'$, we can write
\begin{align*}
&\Omega(\B{\tilde{\chi}}{i'}{k'}{j'}{y',l'}{x}; x',\cdots,
x_r)=\Omega(\B{\tilde{\chi}}{i'}{k'}{j'}{y',l'}{x}; x')\times_{x_\alpha\in
N_{xx'}}\Omega(\B{\tilde{\chi}}{i'}{k'}{j'}{y',l'}{x}; x'; x_\alpha),
\end{align*}
such that 
$$
\left\lbrace \begin{array}{ll}
\Omega(\B{R}{\rcv{x'}/i'}{k'}{j'}{y',l'}{x}; x')&=\{\mathcal{H}_{x'}(t_n)|\underbrace{(\chi,j)_{x'} = (R_{\snt{x}},j')}_{\textit{$j'$-th step of sending RTS to $x$}}\}\\
\Omega(\B{R}{\ovh{x'}/i'}{k'}{j'}{y',l'}{x}; x')&=\{\mathcal{H}_{x'}(t_n)|\underbrace{(\chi,j)_{x'} = (R_{\snt{z}},j'),z\ne x}_{\textit{$j'$-th step of sending RTS not to $x$}}\}\\
\Omega(\B{C}{\ovh{x'}/i'}{k'}{j'}{y',l'}{x}; x')&=\{\mathcal{H}_{x'}(t_n)|\underbrace{(\chi,j)_{x'} = (C_{\snt{z}},j'),z\notin N_x}_{\textit{$j'$-th step of sending CTS not to $x$ and $N_x$}}\}\\
\Omega(\B{C}{\rcv{x'}/i'}{k'}{j'}{y',l'}{x}; x')&=\{\mathcal{H}_{x'}(t_n)|\underbrace{(\chi,j)_{x'} = (C_{\snt{x}},j')}_{\textit{$j'$-th step of sending CTS to $x$}}\}\\
\Omega(\B{A}{\rcv{x'}/i'}{k'}{j'}{y',l'}{x}; x')&=\{\mathcal{H}_{x'}(t_n)|\underbrace{(\chi,j)_{x'} = (A_{\snt{x}},j')}_{\textit{$j'$-th step of sending DATA to $x$}}\}
\end{array} \right. ,
$$
where $\Omega(\B{\tilde{\chi}}{i'}{k'}{j'}{y',l'}{x}; x'; x_\alpha)$ contains
all the possible states of neighboring node $x_\alpha$ in $N_{xx'}$ given
ongoing communication between $x$ and $x'$. If $x$ is receiving a RTS or
overhearing a RTS/CTS from $x'$ ($\tilde{\chi} \in \{R_{\rcv{x'}}, R_{\ovh{x'}},
C_{\ovh{x'}}\}$), then for the hidden nodes $x_\alpha$, $x$ should appear to be
in a back-off or idle state since the conversation between $x$ and $x'$ are
concealed:
\begin{equation*}
\Omega(\B{\tilde{\chi}}{i'}{k'}{j'}{y',l'}{x}; x';
x_\alpha)=\Omega(\B{B}{i'}{k'}{0}{y',l'}{x}; x_\alpha), \qquad
\tilde{\chi}\in\{R_{\rcv{x'}}, R_{\ovh{x'}}, C_{\ovh{x'}}\} .\\
\end{equation*}
On the other hand, if $x$ is receiving a CTS or DATA from $x'$, then $x_\alpha$
should be informed because of the network allocation vector (NAV) incorporated
inside the previous RTS/CTS packets sent from $x$. As a result,  $x_\alpha$
should be in corresponding step NAV delay. If not, it is also impossible for
$x_\alpha$ to receive any CTS/DATA packets because its own RTS/CTS handshakes
should have failed. To summarize, we have the following:
\begin{multline*}
\Omega(\B{C}{\rcv{x'}/i'}{k'}{j'}{y',l'}{x}; x'; x_\alpha)=\Omega(\B{B}{i'}{k'}{0}{y',l'}{x}; x_\alpha)\bigcup \{\mathcal{H}_{x'}(t_n)|\\
\underbrace{(\chi, j)_{x_\alpha}=(D_{x}, t_{NAVc}+1+j')}_{\textit{$t_{NAVc}+1+j'$-th step of NAV delay}}\ \& \ \underbrace{\chi_{x_\alpha}\notin \{C_{\rcv{z}}, A_{\rcv{z}}\}}_{\textit{not receiving CTS/DATA}}\ \& \ \underbrace{(\chi, j)_{x_\alpha} \ne (R_{\rcv{z}}, 0)}_{\textit{not begin to send CTS}}\}
\end{multline*}
\begin{multline*}
\Omega(\B{A}{\rcv{x'}/i'}{k'}{j'}{y',l'}{x}; x'; x_\alpha)=\Omega(\B{B}{i'}{k'}{0}{y',l'}{x}; x_\alpha)\bigcup \{\mathcal{H}_{x'}(t_n)|\\
\underbrace{(\chi, j)_{x_\alpha}=(D_{x}, j')}_{\textit{$j'$-th step of NAV
delay}}\ \& \ \underbrace{\chi_{x_\alpha}\notin \{C_{\rcv{z}},
A_{\rcv{z}}\}}_{\textit{not receiving CTS/DATA}}\} .
\end{multline*}
Note that in general $t_{DATA}\gg t_{RTS}(t_{CTS}) $, thus it is possible that
$x_\alpha$ finishes receiving a RTS and starts to broadcast a CTS during the
period of DATA reception at $x$.

At the next time step $t_{n+1}$, $x$ will continue to receive from $x'$ unless some neighbors starts to broadcast, thus
\begin{multline*}
 \Omega_{2}(\B{R}{\rcv{x'}/i'}{k'}{j'}{y',l'}{x}; x', \cdots, x_r) =\Omega(\B{R}{\rcv{x'}/i'}{k'}{j'}{y',l'}{x}; x')\times_{x_\alpha\in N_{xx'}}\\
 \{\mathcal{H}_{x_\alpha}(t_n)\in \Omega(\B{R}{\rcv{x'}/i'}{k'}{j'}{y',l'}{x};x'; x_\alpha)|\underbrace{(\chi, k, j)_{x_\alpha} \notin\{(B, 0, 0),(R_{\rcv{z}},k, 0),(C_{\rcv{z}}, 0, 0)\}}_{\textit{not begin to send RTS/CTS/DATA}}\}
\end{multline*}
\begin{multline*}
 \Omega_{3}(\B{R}{\ovh{x'}/i'}{k'}{j'}{y',l'}{x}; x', \cdots, x_r) =\Omega(\B{R}{\ovh{x'}/i'}{k'}{j'}{y',l'}{x}; x')\times_{x_\alpha\in N_{xx'}}\\
 \{\mathcal{H}_{x_\alpha}(t_n)\in \Omega(\B{R}{\ovh{x'}/i'}{k'}{j'}{y',l'}{x};x'; x_\alpha)|\underbrace{(\chi, k, j)_{x_\alpha} \notin\{(B, 0, 0),(R_{\rcv{z}},k, 0),(C_{\rcv{z}}, 0, 0)\}}_{\textit{not begin to send RTS/CTS/DATA}}\}
\end{multline*}
\begin{multline*}
 \Omega_{4}(\B{C}{\ovh{x'}/i'}{k'}{j'}{y',l'}{x}; x', \cdots, x_r) =\Omega(\B{C}{\ovh{x'}/i'}{k'}{j'}{y',l'}{x}; x')\times_{x_\alpha\in N_{xx'}}\\
 \{\mathcal{H}_{x_\alpha}(t_n)\in \Omega(\B{C}{\ovh{x'}/i'}{k'}{j'}{y',l'}{x};x'; x_\alpha)|\underbrace{(\chi, k, j)_{x_\alpha} \notin\{(B, 0, 0),(R_{\rcv{z}},k, 0),(C_{\rcv{z}}, 0, 0)\}}_{\textit{not begin to send RTS/CTS/DATA}}\}
\end{multline*}
\begin{multline*}
 \Omega_{6}(\B{A}{\rcv{x'}/i'}{k'}{j'}{y',l'}{x}; x', \cdots, x_r) =\Omega(\B{A}{\rcv{x'}/i'}{k'}{j'}{y',l'}{x}; x')\times_{x_\alpha\in N_{xx'}}\\
 \{\mathcal{H}_{x_\alpha}(t_n)\in \Omega(\B{A}{\rcv{x'}/i'}{k'}{j'}{y',l'}{x};x'; x_\alpha)|\underbrace{(\chi, k, j)_{x_\alpha} \notin\{(B, 0, 0),(R_{\rcv{z}},k, 0)\}}_{\textit{not begin to send RTS/CTS}}\}
\end{multline*}
\begin{multline*}
 \Omega_{10}(\B{C}{\rcv{x'}/i'}{k'}{j'}{y',l'}{x}; x', \cdots, x_r) =\Omega(\B{C}{\rcv{x'}/i'}{k'}{j'}{y',l'}{x}; x')\times_{x_\alpha\in N_{xx'}}\\
 \{\mathcal{H}_{x_\alpha}(t_n)\in \Omega(\B{C}{\rcv{x'}/i'}{k'}{j'}{y',l'}{x};x'; x_\alpha)|\underbrace{(\chi, k, j)_{x_\alpha} \ne (B, 0, 0)}_{\textit{not begin to send RTS}}\}
\end{multline*}
and the transition probability functions during receiving are given by
\begin{align*}
&\circled{$2_a$}=\frac{\mathcal{F}_{\Omega_{2}}(\B{R}{\rcv{x'}/i'}{k'}{j'}{y',l'
}{x})}{\mathcal{F}_{\Omega}(\B{R}{\rcv{x'}/i'}{k'}{j'}{y',l'}{x})}, &
&\circled{$3_a$}=\frac{\mathcal{F}_{\Omega_{3}}(\B{R}{\ovh{x'}/i'}{k'}{j'}{y',l'
}{x})}{\mathcal{F}_{\Omega}(\B{R}{\ovh{x'}/i'}{k'}{j'}{y',l'}{x})}, & & 
\circled{$4_a$}=\frac{\mathcal{F}_{\Omega_{4}}(\B{C}{\ovh{x'}/i'}{k'}{j'}{y',l'}
{x})}{\mathcal{F}_{\Omega}(\B{C}{\ovh{x'}/i'}{k'}{j'}{y',l'}{x})} ,\\
&
\circled{$6_a$}=\frac{\mathcal{F}_{\Omega_{6}}(\B{A}{\rcv{x'}/i'}{k'}{j'}{y',l'}
{x})}{\mathcal{F}_{\Omega}(\B{A}{\rcv{x'}/i'}{k'}{j'}{y',l'}{x})},  &&
\circled{$10_a$}=\frac{\mathcal{F}_{\Omega_{10}}(\B{C}{\rcv{x'}/i'}{k'}{j'}{y',
l'}{x})}{\mathcal{F}_{\Omega}(\B{C}{\rcv{x'}/i'}{k'}{j'}{y',l'}{x})}. &
\end{align*}

\subsubsection{End of sending}
 At the last step of transmissions from $x$ to $x'$, $\mathcal{H}_x(t_n)=(i',k',\tilde{\chi},0,\langle y',l' \rangle)$, $\tilde{\chi}\in\{R_{\snt{x'}}, C_{\snt{x'}}\}$, we know the communications are successful only if $x'$ also reaches the last step of receiving. Thus we only consider the joint state probability functions between $x$ and $x'$ in this case:
\begin{eqnarray*}
\circled{$9_a$} &=&\sum_{\Omega_{9}(\B{R}{\snt{x'}/i'}{0}{0}{y',l'}{x};x')}
P^{(n)}(\B{R}{\snt{x'}/i'}{0}{0}{y',l'}{x}, [\chi_{i,k,j}^{\langle y,l
\rangle}]_{x'})\frac{1}{P^{(n)}([R_{\snt{x'}/i',0,0}^{x',l'}]_{x})} , \\
\circled{$5_a$} &=& \sum_{\Omega_{5}([C_{\snt{x'}/i',k',0}^{\langle y',l'
\rangle}]_{x};x')} P^{(n)}([C_{\snt{x'}/i',k',0}^{\langle y',l' \rangle}]_{x},
[\chi_{i,k,j}^{\langle y,l
\rangle}]_{x'})\frac{1}{P^{(n)}([C_{\snt{x'}/i',k',0}^{\langle y',l'
\rangle}]_{x})} ,
\end{eqnarray*}
where
\begin{align*}
\Omega_{9}(\B{R}{\snt{x'}/i'}{0}{0}{y',l'}{x};x')=\{\mathcal{H}_{x'}
(t_n)|\underbrace{(\chi, j)_{x'} =(R_{\rcv{x}}, 0)}_{\textit{last step of
receiving RTS from $x$}}\} ,\\
\Omega_{5}(\B{C}{\snt{x'}/i'}{k'}{0}{y',l'}{x};x')=\{\mathcal{H}_{x'}
(t_n)|\underbrace{(\chi, j)_{x'} =(C_{\rcv{x}}, 0)}_{\textit{last step of
receiving CTS from $x$}}\} .
\end{align*}
Furthermore, since the RTS/CTS $x'$ received must be sent from $x$, we have
\begin{align*}
P^{(n)}([R_{\rcv{x}/i,k,0}^{\langle y,l
\rangle}]_{x'})=\mathcal{F}_{\Omega}([R_{\rcv{x}/i,k,0}^{\langle y,l
\rangle}]_{x'})=\sum_{\Omega([R_{\rcv{x}/i,k,0}^{\langle y,l
\rangle}]_{x'};x)}P^{(n)}([R_{\rcv{x}/i,k,0}^{\langle y,l \rangle}]_{x'},
\B{\chi}{i}{k}{j}{y,l}{x}), \\
P^{(n)}([C_{\rcv{x}/i,0,0}^{\langle x,l\rangle}]_{x'})=
\mathcal{F}_{\Omega}([C_{\rcv{x}/i,0,0}^{\langle
x,l\rangle}]_{x'})=\sum_{\Omega([C_{\rcv{x}/i,0,0}^{\langle
x,l\rangle}]_{x'};x)}P^{(n)}([C_{\rcv{x}/i,0,0}^{\langle x,l\rangle}]_{x'},
[\chi_{i,k,j}^{\langle y,l \rangle}]_{x}),
\end{align*}
where $i$ and $l$ are fixed for $x'$ and
\begin{align*}
\Omega(\B{R}{\rcv{x}/i}{k}{0}{y,l}{x'};x)=\{\mathcal{H}_{x}(t_n)|\underbrace{
(\chi, j)_{x} =(R_{\snt{x'}}, 0)}_{\textit{last step of sending RTS to $x'$}}\},
\\
\Omega(\B{C}{\rcv{x'}/i'}{0}{0}{x,l}{x'};x)=\{\mathcal{H}_{x}(t_n)|\underbrace{
(\chi, j)_{x} =(C_{\snt{x'}}, 0)}_{\textit{last step of sending CTS to $x'$}}\}.
\end{align*}
We can now rewrite the transition probability functions as
\begin{eqnarray*}
\circled{$9_a$}&=&\sum_{\Omega_{9}([R_{\snt{x'}/i',0,0}^{x',l'}]_{x};x')}
\Bigg(\frac{P^{(n)}([R_{\snt{x'}/i',0,0}^{x',l'}]_{x}, [\chi_{i,k,j}^{\langle
y,l \rangle}]_{x'})}{\mathcal{F}_{\Omega}([R_{\rcv{x}/i,k,0}^{\langle y,l
\rangle}]_{x'})}\frac{P^{(n)}([R_{\rcv{x}/i,k,0}^{\langle y,l
\rangle}]_{x'})}{P^{(n)}([R_{\snt{x'}/i',0,0}^{x',l'}]_{x})}\Bigg) ,\\
\circled{$5_a$} &=& \sum_{\Omega_{5}([C_{\snt{x'}/i',k',0}^{\langle y',l'
\rangle}]_{x};x')}\Bigg(\frac{P^{(n)}([C_{\snt{x'}/i',k',0}^{\langle y',l'
\rangle}]_{x}, [\chi_{i,k,j}^{\langle y,l
\rangle}]_{x'})}{\mathcal{F}_{\Omega}([C_{\rcv{x}/i,0,0}^{\langle
x,l\rangle}]_{x'})}\frac{P^{(n)}([C_{\rcv{x}/i,0,0}^{\langle
x,l\rangle}]_{x'})}{P^{(n)}([C_{\snt{x'}/i',k',0}^{\langle y',l'
\rangle}]_{x})}\Bigg).\\
\end{eqnarray*}

\subsubsection{End of waiting}
 Finally, suppose at time $t_n$ node $x$ is waiting (busy channel/NAV/CTS timeout) as well as sensing the channel: $\mathcal{H}_x(t_n)=(i',k',\tilde{\chi},0,\langle y',l' \rangle)$, $\tilde{\chi}\in\{U,D_{z},W\}$. If a busy channel is sensed by $x$ during backing-off or idle, then there must be at least one active neighbor accessing the channel at the same time. Moreover, it is impossible for $x$ to send any packet, or receive a responding CTS (as $x$ will be a sender in that case). Thus
\begin{align*}
&\Omega(\B{U}{x'/i'}{k'}{0}{y',l'}{x}; x_1,\cdots,x_r)=\times_{x_\alpha\in N_x}\Omega(\B{U}{x'/i'}{k'}{0}{y',l'}{x}; x_\alpha;x_1,\cdots,x_{\alpha-1},x_{\alpha+1},\cdots, x_r)\\
&=\bigcup_{x_\alpha\in N_x}\Omega(\B{U}{x'/i'}{k'}{0}{y',l'}{x}; x_\alpha)\times_{x_\beta\in N_x\backslash x_\alpha}\Omega(\B{U}{x'/i'}{k'}{0}{y',l'}{x}; x_\alpha;x_\beta)\\
&=\bigcup_{x_\alpha\in N_x}\{\mathcal{H}_{x_\alpha}(t_n)|\underbrace{\chi_{x_\alpha} \in \{R_{\snt{z}}, C_{\snt{z'}}, A_{\snt{z}}\},z'\ne x}_{\textit{transmitting (except CTS to $x$)}}\}\\
&\qquad \times_{x_\beta\in N_x\backslash x_\alpha}\{\mathcal{H}_{x_\beta}(t_n)|\underbrace{\chi_{x_\beta} \notin \{R_{\rcv{x}/\ovh{x}}, C_{\rcv{x}/\ovh{x}}, A_{\rcv{x}}\}}_{\textit{not receiving from $x$}}\ \& \ \underbrace{\chi_{x_\beta}\ne C_{\snt{x}}}_{\textit{not sending CTS to $x$}}\}
\end{align*}
At the next time step if at least one neighbor is at the end of transmitting
while no neighbors are in the middle of or ready to initiate a broadcasting, $x$
will sense a free channel again and resume back-off counting (or become idle if
the queue is empty) at the next time step. Otherwise, $x$ will continue waiting.
Therefore,
\begin{align*}
&\Omega_8(\B{U}{x'/i'}{k'}{0}{y',l'}{x}; x_1,\cdots,x_r)\\
&=\bigcup_{x_\alpha\in N_x}\{\mathcal{H}_{x_\alpha}(t_n)|\underbrace{(\chi,j)_{x_\beta} \in \{(R_{\snt{z}},0), (C_{\snt{z'}},0), (A_{\snt{z}},0)\},z'\ne x}_{\textit{end of transmitting (except CTS to $x$)}}\}\\
&\qquad \times_{x_\beta\in N_x\backslash x_\alpha}\{\mathcal{H}_{x_\beta}(t_n)\in \Omega(\B{U}{x'/i'}{k'}{0}{y',l'}{x}; x_\alpha;x_\beta)|\\
&\qquad \qquad \underbrace{(\chi,j)_{x_\beta} \notin \{(R_{\snt{z}},j'), (C_{\snt{z'}},j'), (A_{\snt{z}},j')\}, j'\ne 0}_{\textit{not in the middle of sending}}\\
&\qquad \qquad \qquad \& \ \underbrace{(\chi, k, j)_{x_\beta} \notin\{(B, 0, 0),(R_{\rcv{z}},k, 0),(C_{\rcv{z}}, 0, 0)\}}_{\textit{not begin to send RTS/CTS/DATA}}\}
\end{align*}
so that
\begin{eqnarray*}
\circled{$8_a$}=\frac{\mathcal{F}_{\Omega_{8}}(\B{U}{i'}{k'}{0}{y',l'}{x})}{\mathcal{F}_{\Omega}(\B{U}{i'}{k'}{0}{y',l'}{x})}
\end{eqnarray*}
If $x$ is at the end of NAV waiting period due to RTS or CTS from $x'$. Using the facts that $x$ can not interact with other nodes during waiting and if $x'$ is sending or receiving DATA, it must be at the last step, we conclude that
\begin{align*}
&\Omega(\B{D}{x'/i'}{k'}{0}{y',l'}{x}; x',\cdots,x_r)\\
&=\{\mathcal{H}_{x'}(t_n)|\underbrace{(\chi,j)_{x'}\notin\{(A_{\snt{z}/\rcv{z}},j')\}, j'\ne 0}_{\textit{not in the middle of DATA}}\ \& \ \underbrace{\chi_{x'}\notin\{R_{\rcv{x}/\ovh{x}}, C_{\snt{x}/\rcv{x}/\ovh{x}}, A_{\snt{x}/\rcv{x}}, D_{x}\}}_{\textit{not interacting with $x$}}\}\\
&\qquad \times_{x_\alpha\in N_x\backslash x'}\{\mathcal{H}_{x_\alpha}(t_n)|\underbrace{\chi_{x_\alpha}\notin\{R_{\rcv{x}/\ovh{x}}, C_{\snt{x}/\rcv{x}/\ovh{x}}, A_{\snt{x}/\rcv{x}}, D_{x}\}}_{\textit{not interacting with $x$}}\}
\end{align*}

At the next time step if all neighbors of $x$ are not transmitting or begin to send, $x$ will detect a free channel and consequentially resume back-off counting (or become idle). Otherwise we assume $x$ will wait until the channel is clear. The corresponding conditions are
\begin{align*}
&\Omega_{7}(\B{D}{x'/i'}{k'}{0}{y',l'}{x}; x', \cdots, x_{r})=\Omega(\B{D}{x'/i'}{k'}{0}{y',l'}{x}; x',\cdots,x_r)\bigcap\times_{x_\alpha\in N_x}\{\mathcal{H}_{x_\alpha}(t_n)|\\
&\qquad \underbrace{\chi_{x_\alpha}\notin\{R_{\snt{z}}, C_{\snt{z}}, A_{\snt{z}}\}}_{\textit{not transmitting}}\ \& \ \underbrace{(\chi, k, j)_{x_\alpha} \notin\{(B, 0, 0),(R_{\rcv{z}},k, 0),(C_{\rcv{z}}, 0, 0)\}}_{\textit{not begin to send RTS/CTS/DATA}}\}
\end{align*}
such that
\begin{eqnarray*}
\circled{$7_a$}=\frac{\mathcal{F}_{\Omega_{7}}(\B{D}{x'/i'}{k'}{0}{y',l'}{x})}{\mathcal{F}_{\Omega}(\B{D}{x'/i'}{k'}{0}{y',l'}{x})}
\end{eqnarray*}
If $x$ is at the last step of CTS timeout for $x'$, which implies the previous RTS/CTS between $x$ and $x'$ fails, the possible concurrent states of $x'$ are
\begin{align*}
&\Omega(\B{W}{i'}{0}{0}{x',l'}{x}; x')=\{\mathcal{H}_{x'}(t_n)|\chi_{x'}\in\{I,B,U,R_{\snt{z}},R_{\rcv{z'}/\ovh{z'}}\}, z'\ne x\}
\end{align*}
For any other neighbor $x_{\alpha}\in N_{x}\backslash x'$, the previous RTS from $x$ maybe overheard. Then
\begin{align*}
&\Omega(\B{W}{i'}{0}{0}{x',l'}{x}; x';x_\alpha)=\Omega(\B{W}{i'}{0}{0}{x',l'}{x}; x_\alpha) \bigcup \{\mathcal{H}_{x_\alpha}(t_n)|\underbrace{(\chi,j)_{x_\alpha}=(D_{x},t_{NAVc})}_{\textit{$t_{NAVc}$-th step of NAV for $x$}}\}
\end{align*}
Thus for all neighboring nodes of $x$ we have
\begin{align*}
\Omega(\B{W}{i'}{0}{0}{x',l'}{x}; x',\cdots,x_r)=\Omega(\B{W}{i'}{0}{0}{x',l'}{x}; x')\times_{x_\alpha\in N_x\backslash x'}\Omega(\B{W}{i'}{0}{0}{x',l'}{x}; x';x_\alpha)
\end{align*}
At the next time step, if no neighbors are sending or begin to send RTS:
\begin{multline*}
\Omega_{11}(\B{W}{i'}{0}{0}{x',l'}{x}; x',\cdots,x_r) = \Omega(\B{W}{i'}{0}{0}{x',l'}{x}; x',\cdots,x_r)\bigcap\\ \times_{x_\alpha\in N_x} \{\mathcal{H}_{x_\alpha}(t_n)|\underbrace{(\chi,k)_{x_\alpha}\ne (B, 0)\ \& \ \chi_{x_\alpha}\ne R_{\snt{z}}}_{\textit{not sending or begin to send RTS}}\}
\end{multline*}
$x$ resumes idle or back-off with probability evaluated by
\begin{eqnarray*}
\circled{$11_a$}=\frac{\mathcal{F}_{\Omega_{11}}(\B{W}{i'}{0}{0}{x',l'}{x})}{\mathcal{F}_{\Omega}(\B{W}{i'}{0}{0}{x',l'}{x})}
\end{eqnarray*}

\subsection{Equilibrium Distribution}
In this section we will set up the balance equations for solving the stationary distribution, $\pi\GB{x}$, of the discrete time Markov chain as $n\to\infty$. i.e. $\pi\GB{x}=\lim_{n\to\infty}P^{(n)}(\GB{x})$.
For the transition probabilities such as $\circled{1a}$, we adopt the following notation:
\begin{equation*}
p^{x}_{1a}=\lim_{n\to \infty}P^{x}_{1a}(t_n):=\lim_{n\to \infty}\circled{$1_a$}
\end{equation*}
If the transition such as $\circled{2a}$ involves a specific neighboring node $x'$, we use
\begin{equation*}
p^{xx'}_{2a}=\lim_{n\to \infty}P^{xx'}_{2a}(t_n):=\lim_{n\to \infty}\circled{$2_a$}
\end{equation*}
\subsubsection{System formulation}
We start building the system from the base layer where $l=0$:
\begin{eqnarray}\label{steadyequ-I}
\pi\B{I}{0}{0}{0}{\emptyset,0}{x}&=&p^x_{1a} \pi\B{I}{0}{0}{0}{\emptyset,0}{x}+p^{x}_{8a} \pi\B{U}{0}{0}{0}{\emptyset,0}{x}+\sum_{z\in N_x}(p^{xz}_{5b} \pi\B{C}{\snt{z}/0}{0}{0}{\emptyset,0}{x}+p^{xz}_{7a} \pi\B{D}{z/0}{0}{0}{\emptyset,0}{x}) \notag \\
&&+\begin{cases}
     \sum_{z\in N_x}(\pi\B{W}{m}{0}{0}{z,1}{x}+\pi\B{A}{\snt{z}/0}{0}{0}{z,1}{x} ),&  L_x >0\\
     \sum_{z\in N_x}\pi\B{A}{\rcv{z}/0}{0}{0}{\emptyset,0}{x}, &  L_x =0
\end{cases}
\end{eqnarray}
$L_x=0$ implies that node $x$ has an empty queue. Next, suppose the queue is
non-empty ( $l>0$) and node $x$ is backing off for node $y$ where $y\in
N_x$. If $k=0$, $x$ transmits immediately, thus 
\begin{equation}
\pi \B{B}{i}{0}{0}{y,l}{x} = p^x_{1a} \pi\B{B}{i}{1}{0}{y,l}{x}\label{steadyequ-B2}
\end{equation}
Otherwise, depending on the value of $i$ (back-off stage), $k$ (back-off counter) and $l$ (size of queue), we have
\begin{eqnarray}\label{steadyequ-B}
\pi \B{B}{i}{k}{0}{y,l}{x}&=&
\begin{cases}
   p^x_{1a} \pi\B{B}{i}{k+1}{0}{y,l}{x} + p^{x}_{8a} \pi\B{U}{i}{k}{0}{y,l}{x}+\sum_{z\in N_x}(p^{xz}_{5b} \pi\B{C}{\snt{z}/i}{k}{0}{y,l}{x}+p^{xz}_{7a} \pi\B{D}{z/i}{k}{0}{y,l}{x}), & 0<k<2^iw \\
      p^{x}_{8a} \pi\B{U}{i}{k}{0}{y,l}{x}+\sum_{z\in N_x}(p^{xz}_{5b} \pi\B{C}{\snt{z}/i}{k}{0}{y,l}{x}+p^{xz}_{7a} \pi\B{D}{z/i}{k}{0}{y,l}{x}), & k=2^iw
\end{cases} \notag\\
&&+\begin{cases}
   \sum_{z\in N_x} \pi\B{A}{\rcv{z}/i}{k}{0}{y,l-1}{x}, & 1<l<L_x \\
   \sum_{z\in N_x} (\pi\B{A}{\rcv{z}/i}{k}{0}{y,L_x-1}{x} +  \pi\B{A}{\rcv{z}/i}{k}{0}{y,L_x}{x}),   &  l=L_x \\
   0, &  l=1
\end{cases}\notag\\
&&+\begin{cases}
  \frac{1}{2^iw} \pi\B{W}{i-1}{0}{0}{y,l}{x}, & i>0 \\
     \sum_{z\in N_x}\frac{P_{xy}}{w} (\pi\B{W}{m}{0}{0}{z,l+1}{x}+\pi\B{A}{\snt{z}/0}{0}{0}{z,l+1}{x}), &  i=0
\end{cases}\notag\\
&&+\begin{cases}
 \sum_{z\in N_x}\frac{P_{xy}}{w} \pi\B{A}{\rcv{z}/0}{0}{0}{\emptyset,0}{x},   &  i=0,\ l=1 \\
   0, &  \text{otherwise}
\end{cases}
\end{eqnarray}
where $P_{xy}$ denotes the probability of $x$ sending a data packet to its
neighbor $y$. 
Within each CSB and RCB, the steady state distribution of $x$ should satisfy:
\begin{eqnarray}\label{steadyequ-A*}
\pi\B{A}{\rcv{z}/i}{k}{0}{y,l}{x}&=\cdots=&(p^{xz}_{6a})^{t_{DATA}}\pi\B{A}{\rcv{z}/i}{k}{t_{DATA}}{y,l}{zx}=(p^{x}_{6a})^{t_{DATA}}\cdot p^{xz}_{5a} \pi\B{C}{\snt{z}/ i}{k}{0}{y,l}{x}\notag\\
&=\cdots=& (p^{xz}_{6a})^{t_{DATA}}\cdot p^{xz}_{5a}\cdot (p^{xz}_{2a})^{t_{RTS}}\pi\B{R}{\rcv{z}/i}{k}{t_{RTS}}{y,l}{x}\notag\\
&=& \big((p^{xz}_{6a})^{t_{DATA}}\cdot p^{xz}_{5a}\cdot (p^{xz}_{2a})^{t_{RTS}}\cdot p^{xz}_{1b}\big) \pi\B{B}{i}{k}{0}{y,l}{x}
\end{eqnarray}

\begin{eqnarray}\label{steadyequ-D}
p^{xz}_{7a} \pi\B{D}{z/i}{k}{0}{y,l}{x}&=\cdots=&\pi\B{D}{z/i}{k}{t_{NAVc}+1}{y,l}{x}+\pi\B{C}{\ovh{z}/i}{k}{0}{y,l}{x}\notag\\
&=\cdots=&(p^{xz}_{3a})^{t_{RTS}} \pi\B{R}{\ovh{z}/i}{k}{t_{RTS}}{y,l}{x}+(p^{xz}_{4a})^{t_{CTS}} \pi\B{C}{\ovh{z}/i}{k}{t_{CTS}}{y,l}{x}\notag\\
&=&\big((p^{xz}_{3a})^{t_{RTS}}\cdot  p^{xz}_{1c}+(p^{xz}_{4a})^{t_{CTS}}\cdot  p^{xz}_{1d}\big)\pi\B{B}{i}{k}{0}{y,l}{x}
\end{eqnarray}

\begin{eqnarray}\label{steadyequ-U}
p^{x}_{8a} \pi\B{U}{i}{k}{0}{y,l}{x}&=&\sum_{z\in N_x}\big(\sum_{j=1}^{t_{RTS}}p^{xz}_{2b} \pi\B{R}{\rcv{z}/i}{k}{j}{y,l}{x}+\sum_{j=1}^{t_{DATA}}p^{xz}_{6b} \pi\B{A}{\rcv{z}/i}{k}{j}{y,l}{x}+\sum_{j=1}^{t_{RTS}}p^{xz}_{3b} \pi\B{R}{\ovh{z}/i}{k}{j}{y,l}{x}\notag\\
&+&\sum_{j=1}^{t_{CTS}}p^{xz}_{4b} \pi\B{C}{\ovh{z}/i}{k}{j}{y,l}{x}\big)+ p^{x}_{1e} \pi\B{B}{i}{k}{0}{y,l}{x}
\end{eqnarray}

\begin{eqnarray}\label{steadyequ-A}
\pi\B{A}{\snt{y}/0}{0}{0}{y,l}{x}&= \cdots =&\sum_{i=0}^{m}(p^{xy}_{10a})^{t_{CTS}}\pi\B{C}{\rcv{y}/0}{0}{t_{CTS}}{y,l}{x}=\sum_{i=0}^{m}(p^{xy}_{10a})^{t_{CTS}}\cdot p^{xy}_{9a}\pi\B{R}{\snt{y}/i}{0}{0}{y,l}{x}\notag\\
&=\cdots =&\sum_{i=0}^{m}(p^{xy}_{10a})^{t_{CTS}}\cdot p^{xy}_{9a} \pi\B{B}{i}{0}{0}{y,l}{x}
\end{eqnarray}

\begin{eqnarray}\label{steadyequ-W}
p^{xy}_{11a}\pi\B{W}{i}{0}{0}{y,l}{x} &=\cdots=&\sum_{j=1}^{t_{CTS}}p^{xy}_{10b} \pi\B{C}{i}{0}{j}{y,l}{x}+\pi\B{W}{i}{0}{t_{out}}{y,l}{x}\notag\\
&=\cdots=& \big(1-(p^{xy}_{10a})^{t_{CTS}}\cdot p^{xy}_{9a}\big)\pi\B{B}{i}{0}{0}{y,l}{x}
\end{eqnarray}
Now using equations (\ref{steadyequ-A*})--(\ref{steadyequ-W}), we can rewrite (\ref{steadyequ-I}) and (\ref{steadyequ-B}) as: 
\begin{align*}
\pi\B{I}{0}{0}{0}{\emptyset,0}{x}&=\big(p^{x}_{1a}+p^{x}_{1e}+\sum_{z\in N_x}(p^{xz}_{1b}(1-P^{xz}_R)+p^{xz}_{1c}+p^{xz}_{1d})\big)\pi\B{I}{0}{0}{0}{\emptyset,0}{x}\\
&\quad +\sum_{z\in N_x}(P^{xz}_S\sum_{i=0}^{m-1} \pi\B{B}{i}{0}{0}{z,1}{x}+\pi\B{B}{m}{0}{0}{z,1}{x})\tag{3.4.1a}
\end{align*}
\begin{align*}\label{steadyequ-Bs}
\pi\B{B}{i}{k}{0}{y,l}{x}&=
\begin{cases}
 	p^{x}_{1a} \pi\B{B}{i}{k+1}{0}{y,l}{x}+\big(p^{x}_{1e}+\sum_{z\in N_x}(p^{xz}_{1b}(1-P^{xz}_R)+p^{xz}_{1c}+p^{xz}_{1d})\big) \pi\B{B}{i}{k}{0}{y,l}{x}, & 1<k<2^iw\\
	\big(p^{x}_{1e}+\sum_{z\in N_x}(p^{xz}_{1b}(1-P^{xz}_R)+p^{xz}_{1c}+p^{xz}_{1d})\big) \pi\B{B}{i}{k}{0}{y,l}{x}, & k=2^iw
	\end{cases}\\
& \quad +\begin{cases}
	\sum_{z\in N_x}p^{xz}_{1b}P^{xz}_R \pi\B{B}{i}{k}{0}{y,l-1}{x}, & 1<l<L_x\\
	\sum_{z\in N_x}p^{xz}_{1b}P^{xz}_R (\pi\B{B}{i}{k}{0}{y,L_x-1}{x}+ \pi\B{B}{i}{k}{0}{y,L_x}{x}), & l=L_x\\
	0 & \text{otherwise}
	\end{cases}\\
 &\quad +\begin{cases}
	\frac{1-P^{xy}_S}{2^iw} \pi\B{B}{i-1}{0}{0}{y,l}{x}, & i>0\\
	\frac{P_{xy}}{w}\sum_{z\in N_x}(P^{xz}_S\sum_{i=0}^{m-1} \pi\B{B}{i}{0}{0}{z,l+1}{x}+\pi\B{B}{m}{0}{0}{z,l+1}{x}), & i=0\\
	\end{cases}\\
&\quad +\begin{cases}
 \sum_{z\in N_x}\frac{P_{xy}}{w}p^{xz}_{1b}P^{xz}_R \pi\B{I}{0}{0}{0}{\emptyset,0}{x},   &  i=0,\ l=1 \\
   0, &  \text{otherwise}
   \end{cases} \tag{3.4.3a}
\end{align*}
where $P^{xz}_S:=(P^{xz}_{10a})^{t_{CTS}}\cdot P^{xz}_{9a}$ represents the probability of a successful sending of a data packet at $x$ to $z$, while $P^{xz}_R:=(p^{xz}_{6a})^{t_{DATA}}\cdot p^{xz}_{5a}\cdot (p^{xz}_{2a})^{t_{RTS}}$ denotes the probability of a successful receiving of a data packet at $x$ from $z$.

%
%

\subsubsection{System Closure} 
Notice that the transition probability functions are still related to unknown joints probability functions. As a first step to conclude a solution, we complete the nonlinear system by applying naive product approximations:
$$
P^{(n)}(\B{\chi}{i}{k}{j}{y,l}{x_1},\B{\chi}{i}{k}{j}{y,l}{x_2}, \dots, \B{\chi}{i}{k}{j}{y,l}{x_r})\approx  \prod_{\alpha=1}^{\alpha=r}P^{(n)}(\B{\chi}{i}{k}{j}{y,l}{x_\alpha})
$$
For simplicity we shall write
$$
\frac{\sum_{\Omega_{\alpha}}\prod_{\gamma=1}^{\gamma=r}P^{(n)}(\GB{{x_\gamma}})}{\sum_{\Omega_{\beta}} \prod_{\gamma=1}^{\gamma=r}P^{(n)}(\GB{{x_\gamma}})}=\mathcal{P}_{\frac{x_1\times\cdots\times x_r}{x_1\times\cdots\times x_r}}\Bigg(\frac{{\Omega_\alpha}}{{\Omega_\beta}}\Bigg)
$$
for any summation conditions $\Omega_\alpha$ and $\Omega_\beta$, and nodes $x_1, \cdots, x_r$. Furthermore, if $\Omega_\alpha$ and $\Omega_\beta$ can be decomposed as $\prod_{\gamma=1}^{\gamma=r}\Omega_{\alpha}(x_\gamma)$ and $\prod_{\gamma=1}^{\gamma=r}\Omega_{\beta}(x_\gamma)$, we can interchange the summation and product and denote:
 $$
 \prod_{\gamma=1}^{\gamma=r}\frac{\sum_{\Omega_{\alpha}}P^{(n)}(\GB{{x_\gamma}})}{\sum_{\Omega_{\beta}}P^{(n)}(\GB{{x_\gamma}})}= \prod_{\gamma=1}^{\gamma=r}\mathcal{P}_{\frac{x_\gamma}{x_\gamma}}\Bigg(\frac{{\Omega_\alpha(x_\gamma)}}{{\Omega_\beta(x_\gamma)}}\Bigg)
$$

In terms of each types of actions in section \ref{subsec: represent}, we summarize the approximations using tables \ref{t: TP-approx1} to \ref{t: TP-approx4}. Note that if $\forall x''\in N_{x}$, $N_{x''} = \{x\}$, all messages from $x$ will be successfully admitted by its neighbors. Thus we should have \circled{$6_a$}$ \approx 1$ and \circled{$10_a$}$ \approx 1$ since the channel has already been reserved by $x$ through NAV.
\begin{table}[!h]
\footnotesize
\caption{\label{t: TP-approx1}
    Transition probability function approximations - case 1}
    \renewcommand{\arraystretch}{1.6}
		\centering
       \begin{tabular}{| c || c |}
       \hline
       Transition Probability & Approximations\\
       \hline
         $P^{x}_{1a}$   &   $\prod_{x_\alpha\in N_x}\mathcal{P}_{\frac{x_\alpha}{x_\alpha}}\Bigg(\frac{{\Omega_{1a}(\B{B}{i'}{k'}{0}{y',l'}{x}; x_\alpha)}}{{\Omega(\B{B}{i'}{k'}{0}{y',l'}{x}; x_\alpha)}}\Bigg)$\\
       \hline
       $P^{xx'}_{1b} $  &   $\mathcal{P}_{\frac{x'}{x'}}\Bigg(\frac{{\Omega_{1b}(\B{B}{i'}{k'}{0}{y',l'}{x}; x')}}{{\Omega(\B{B}{i'}{k'}{0}{y',l'}{x}; x')}}\Bigg)\prod_{x_\alpha\in N_x\backslash x'}\mathcal{P}_{\frac{x_\alpha}{x_\alpha}}\Bigg(\frac{{\Omega_{1a}(\B{B}{i'}{k'}{0}{y',l'}{x}; x_\alpha)}}{{\Omega(\B{B}{i'}{k'}{0}{y',l'}{x}; x_\alpha)}}\Bigg)$\\
       \hline
       $P^{xx'}_{1c} $  &   $\mathcal{P}_{\frac{x'}{x'}}\Bigg(\frac{{\Omega_{1c}(\B{B}{i'}{k'}{0}{y',l'}{x}; x')}}{{\Omega(\B{B}{i'}{k'}{0}{y',l'}{x}; x')}}\Bigg)\prod_{x_\alpha\in N_x\backslash x'}\mathcal{P}_{\frac{x_\alpha}{x_\alpha}}\Bigg(\frac{{\Omega_{1a}(\B{B}{i'}{k'}{0}{y',l'}{x}; x_\alpha)}}{{\Omega(\B{B}{i'}{k'}{0}{y',l'}{x}; x_\alpha)}}\Bigg)$\\
       \hline
       $P^{xx'}_{1d} $  &   $\mathcal{P}_{\frac{x'}{x'}}\Bigg(\frac{{\Omega_{1d}(\B{B}{i'}{k'}{0}{y',l'}{x}; x')}}{{\Omega(\B{B}{i'}{k'}{0}{y',l'}{x}; x')}}\Bigg)\prod_{x_\alpha\in N_x\backslash x'}\mathcal{P}_{\frac{x_\alpha}{x_\alpha}}\Bigg(\frac{{\Omega_{1a}(\B{B}{i'}{k'}{0}{y',l'}{x}; x_\alpha)}}{{\Omega(\B{B}{i'}{k'}{0}{y',l'}{x}; x_\alpha)}}\Bigg)$\\
       \hline
       $P^{x}_{1e} $  &  $1-P^{x}_{1a}-\sum_{x'\in N_x}(P^{xx'}_{1b}+P^{xx'}_{1c}+P^{xx'}_{1d})$\\
       \hline
       \end{tabular}
\end{table}

\begin{table}[!h]
\footnotesize
\caption{\label{t: TP-approx2}
    Transition probability function approximations - case 2}
    \renewcommand{\arraystretch}{1.6}
		\centering
       \begin{tabular}{| c || c |}
       \hline
       Transition Probability & Approximations\\
       \hline
         $P^{x}_{2a}$   &   $\prod_{x_\alpha\in N_{xx'}}\mathcal{P}_{\frac{x_\alpha}{x_\alpha}}\Bigg(\frac{{\Omega_{2}(\B{R}{\rcv{x'}/i'}{k'}{j'}{y',l'}{x};x'; x_\alpha)}}{{\Omega(\B{B}{i'}{k'}{0}{y',l'}{x}; x_\alpha)}}\Bigg)$\\
       \hline
       $P^{xx'}_{3a} $  &   $\prod_{x_\alpha\in N_{xx'}}\mathcal{P}_{\frac{x_\alpha}{x_\alpha}}\Bigg(\frac{{\Omega_{3}(\B{R}{\ovh{x'}/i'}{k'}{j'}{y',l'}{x};x'; x_\alpha)}}{{\Omega(\B{B}{i'}{k'}{0}{y',l'}{x}; x_\alpha)}}\Bigg)$\\
       \hline
       $P^{xx'}_{4a} $  &  $\prod_{x_\alpha\in N_{xx'}}\mathcal{P}_{\frac{x_\alpha}{x_\alpha}}\Bigg(\frac{{\Omega_{4}(\B{C}{\ovh{x'}/i'}{k'}{j'}{y',l'}{x};x'; x_\alpha)}}{{\Omega(\B{B}{i'}{k'}{0}{y',l'}{x}; x_\alpha)}}\Bigg)$\\
       \hline
       $P^{xx'}_{6a} $  &  $\prod_{x_\alpha\in N_{xx'}}\mathcal{P}_{\frac{x_\alpha}{x_\alpha}}\Bigg(\frac{{\Omega_{6}(\B{A}{\rcv{x'}/i'}{k'}{j'}{y',l'}{x};x'; x_\alpha)}}{{\Omega(\B{A}{\rcv{x'}/i'}{k'}{j'}{y',l'}{x};x'; x_\alpha)}}\Bigg)$\\
       \hline
       $P^{xx'}_{10a} $  & $\prod_{x_\alpha\in N_{xx'}}\mathcal{P}_{\frac{x_\alpha}{x_\alpha}}\Bigg(\frac{{\Omega_{10}(\B{C}{\rcv{x'}/i'}{k'}{j'}{y',l'}{x};x'; x_\alpha)}}{{\Omega(\B{C}{\rcv{x'}/i'}{k'}{j'}{y',l'}{x};x'; x_\alpha)}}\Bigg)$\\
       \hline
       \end{tabular}
\end{table}

\begin{table}[!h]
\footnotesize
\caption{\label{t: TP-approx3}
    Transition probability function approximations - case 3}
    \renewcommand{\arraystretch}{1.6}
		\centering
       \begin{tabular}{| c || c |}
       \hline
       Transition Probability & Approximations\\
       \hline
         $P^{xx'}_{5a}$   &   $\mathcal{P}_{\frac{x'}{x}}\Bigg(\frac{{\Omega_{5}(\B{C}{\snt{x'}/i'}{k'}{0}{y',l'}{x};x')}}{{\Omega(\B{C}{\rcv{x}/i}{0}{0}{x,l}{x'}; x)}}\Bigg)$\\
       \hline
       $P^{xx'}_{9a} $  &  $\mathcal{P}_{\frac{x'}{x}}\Bigg(\frac{{\Omega_{9}(\B{R}{\snt{x'}/i'}{0}{0}{x',l'}{x};x')}}{{\Omega(\B{R}{\rcv{x}/i}{k}{0}{y,l}{x'}; x)}}\Bigg)$\\
       \hline
       \end{tabular}
\end{table}

\begin{table}[!h]
\footnotesize
\caption{\label{t: TP-approx4}
    Transition probability function approximations - case 4}
    \renewcommand{\arraystretch}{1.6}
		\centering
       \begin{tabular}{| c || c |}
       \hline
       Transition Probability & Approximations\\
       \hline
         $P^{xx'}_{7a}$   &   $\mathcal{P}_{\frac{x' \times \cdots\times x_r}{x' \times \cdots\times x_r}}\Bigg(\frac{\Omega_{7}(\B{D}{x'/i'}{k'}{j'}{y',l'}{x};x',\cdots, x_r)}{\Omega(\B{D}{x'/i'}{k'}{j'}{y',l'}{x};x',\cdots, x_r)}\Bigg)$\\
       \hline
       $P^{x}_{8a} $  &  $\mathcal{P}_{\frac{x_1 \times \cdots\times x_r}{x_1 \times \cdots\times x_r}}\Bigg(\frac{\Omega_{8}(\B{U}{i'}{k'}{j'}{y',l'}{x};x_1,\cdots, x_r)}{\Omega(\B{U}{i'}{k'}{j'}{y',l'}{x};x_1,\cdots, x_r)}\Bigg)$\\
       \hline
       $P^{xx'}_{11a} $  & $\mathcal{P}_{\frac{x' \times \cdots\times x_r}{x' \times \cdots\times x_r}}\Bigg(\frac{\Omega_{11}(\B{W}{i'}{0}{0}{x',l'}{x};x',\cdots, x_r)}{\Omega(\B{D}{i'}{0}{0}{x',l'}{x};x',\cdots, x_r)}\Bigg)$\\
       \hline
       \end{tabular}
\end{table}

\section{Examples}
\label{sec: example}
\subsection{QualNet Simulation}
To validate our model, we compare the detailed equilibrium node
states of three representative networks 
with results from realistic simulations of wireless networks in the QualNet
simulator \cite{Qual2010} . QualNet employs high fidelity wireless channel models and
model of IEEE 802.11 DCF.
The parameters used in the QualNet 5.0
simulator are summarized in Table \ref{t:qualnet-params}. Accordingly, the model
parameters used in all examples are concluded as follows. (i) The RTS retry
limit $m$ varies from $0$ to $\sqrt{\frac{\text{CWmax}}{\text{CWmin}}}=2$, that
is, we allow RTS to retransmit $0$, $1$ or $2$ times in each example. (ii) The
initial window size $w$ is set as $\text{CWmin} = 3$. (iii) The transmission
time of RTS/CTS are discretized as $t_{RTS}=t_{CTS}=\left
\lceil{\frac{T_{RTS}}{\sigma}}\right \rceil -1 =1$, similarly $t_{DATA}= 5$ and
$t_{out}=2$. Note that $t_{DATA}$ combines the 
transmission time of both data payload and ACK frame. (iv) The NAV contained in RTS frame should include the remaining time of a complete RTS/CTS/DATA/ACK handshakes thus $t_{NAVr}= \left \lceil{\frac{T_{CTS}+T_{DATA}+T_{ACK}}{\sigma}}\right \rceil -1=7$, similarly $t_{NAVc}=5$.

\begin{table}[!h]
\footnotesize
\caption{\label{t:qualnet-params}
    Parameters used in QualNet simulation}
\centering
       \begin{tabular}{|p{.4\columnwidth}|p{.4\columnwidth}|}
       \hline
       \hline
       Terrain size             & 1500$\times$1500 m$^2$ \\
  Mobility                 & 0\\
  Radio range              & up to 500 m\\
  PHY protocol             & 802.11b\\
  Bandwidth                & 5 Mbps\\
  MAC protocol             & MAC 802.11b\\
  Slot time                     & 140 Microsecond \\ 
  SIFS                             & 0 Microsecond \\ 
  DIFS                             & 140 Microsecond \\ 
  RTS/CTS/ACK Tx time        &    280 Microsecond \\
  CTS Timeout time           &    420 Microsecond \\
  Data Tx time               &    562 Microsecond \\
  CWmin                             & 3 \\ 
  CWmax                            & 12 \\ 
       \hline
       \hline
       \end{tabular}
\end{table}

\subsection{A 2-node network.}
\label{sec: 2-node}
We first consider the following scenario. The network contains two nodes, $x_1$ and $x_2$. Both nodes have infinite number of packets in their queue and consecutively transmit to each other. 
Due to symmetry, we only focus on the behaviors of node $x_1$. Since there is no hidden terminal problem in this simple network and channel conditions are assumed to be ideal, all transmissions are guaranteed against collision and interference unless both nodes reach out at the same moment. 

In this case only four non-trivial transition probabilities exists at any time
step $t_n$, which are described in Table \ref{tab:four}.
\begin{table}[!h]
 \caption{Non-Trivial Transition Probabilities}\label{t: Tp - 2nodes}
\footnotesize
\renewcommand{\arraystretch}{1.4}
\centering
\begin{tabular} {|  c  || p{.4 \columnwidth}  |}   
  \hline
  $\circled{$1_a$}$ & $x_1$ detects a quiet channel while back-off\\ \hline
  $\circled{$1_b$}$&  $x_1$ detects a RTS while back-off \\ \hline
  $\circled{$9_a$}$ & RTS from $x_1$ succeeds\\ \hline
  $\circled{$9_b$}$ & RTS from $x_1$ fails \\ \hline
\end{tabular}
\label{tab:four}
\end{table}

Using product approximations, we evaluate the above transition probabilities as follows in terms of the marginal densities of $x_2$ only. 
\begin{align} \label{TP1a-2nodes}
 P_{1a}^{x_1}(t_n)&=\TP{x_2}{x_2}{\sum_{\chi=B, k\neq 0}+\sum_{\chi=W}}{\sum_{\chi\in\{B,W\}}} \\
P_{1b}^{x_1x_2}(t_n)&= \TP{x_2}{x_2}{\sum_{\chi=B,k=0}}{\sum_{\chi\in\{B,W\}}}\\
P_{9a}^{x_1x_2}(t_n)&=\TP{x_2}{x_1}{\sum_{\chi=R_{\rcv{x_1}},j=0}}{\sum_{\chi=R_{\snt{x_2}},j=0}}\\
P_{9b}^{x_1x_2}(t_n)&=1-P_{9a}^{x_1x_2}(t_n) \label{TP9b-2nodes}
\end{align}
 For the next step, let $t_n\to \infty$ and employing the equilibrium equations
(\ref{steadyequ-B2}), (\ref{steadyequ-Bs}), (\ref{steadyequ-A*}), (\ref{steadyequ-A}) and (\ref{steadyequ-W}), we have: 
 \begin{align}
 \pi\B{B}{i}{0}{0}{x_2,\infty}{x_1}&=p^{x_1}_{1a} \pi\B{B}{i}{1}{0}{x_2,\infty}{x_1}\\
 \pi\B{B}{i}{k}{0}{x_2,\infty}{x_1}&=
 \begin{cases}
  	p^{x_1}_{1a} \pi\B{B}{i}{k+1}{0}{x_2,\infty}{x_1}+p^{x_1x_2}_{1b} \pi\B{B}{i}{k}{0}{x_2,\infty}{x_1}, & 0<k<2^i*3\\
 	p^{x_1x_2}_{1b} \pi\B{B}{i}{2^iw}{0}{x_2,\infty}{x_1}, & k=2^i*3
 \end{cases}\\
  &\quad +\begin{cases}
 	\frac{1}{2^i*3} p_{9b}^{x_1x_2}\pi\B{B}{i-1}{0}{0}{x_2,\infty}{x_1}, & i>0, k\ne 0\\
 	\frac{1}{3}(p^{x_1x_2}_{9a}\sum_{i=0}^{m-1} \pi\B{B}{i}{0}{0}{x_2,\infty}{x_1}+\pi\B{B}{m}{0}{0}{x_2,\infty}{x_1}), & i=0, k\ne 0\\
 	\end{cases} \label{steadyequ: Bs_ex1}.
 \end{align} 
 \begin{align}
 \pi\B{A}{\rcv{x_2}/i}{k}{0}{x_2,\infty}{x_1}&=\cdots=\pi\B{A}{\rcv{x_2}/i}{k}{5}{x_2,\infty}{x_1}= \pi\B{C}{\snt{x_2}/ i}{k}{0}{x_2,\infty}{x_1}=\cdots= \pi\B{R}{\rcv{x_2}/i}{k}{1}{x_2,\infty}{x_1}= p^{x_1x_2}_{1b} \pi\B{B}{i}{k}{0}{x_2,\infty}{x_1}\\
 \pi\B{A}{\snt{x_2}/0}{0}{0}{x_2,\infty}{x_1}&= \cdots =\sum_{i=0}^{m}\pi\B{C}{\rcv{x_2}/i}{0}{1}{x_2,\infty}{x_1}=\sum_{i=0}^{m} p^{x_1x_2}_{9a}\pi\B{R}{\snt{x_2}/i}{0}{0}{x_2,\infty}{x_1}=\cdots =\sum_{i=0}^{m}p^{x_1x_2}_{9a} \pi\B{B}{i}{0}{0}{x_2,\infty}{x_1}\\
 \pi\B{W}{i}{0}{0}{x_2,\infty}{x_1}&=\cdots=\pi\B{W}{i}{0}{2}{x_2,\infty}{x_1}=p^{x_1x_2}_{9b}\pi\B{R}{\snt{x_2}/i}{0}{0}{x_2,\infty}{x_1}=\cdots=p^{x_1x_2}_{9b}\pi\B{B}{i}{0}{0}{x_2,\infty}{x_1}\label{steadyequ: Os_ex1}
 \end{align} 
  Given the transition probability functions from \eqref{TP1a-2nodes} to \eqref{TP9b-2nodes}, together with the symmetry conditions: 
  \begin{equation}
  \pi\B{\chi}{i}{k}{j}{x_2,\infty}{x_1}=\pi\B{\chi}{i}{k}{j}{x_1,\infty}{x_2}
   \end{equation}
  for any $i,k,\chi,j$, and the normalization constraints
  \begin{equation}
\sum_{i,k,\chi,j}\pi\B{\chi}{i}{k}{j}{x_2,\infty}{x_1}=\sum_{i,k,\chi,j}\pi\B{
\chi}{i}{k}{j}{x_1,\infty}{x_2}=1,
  \end{equation}
 the resulting non-linear system can be solved for stationary distributions at
$x_1$ using Matlab's $\mathbf{fsolve}$ subroutine. The initial conditions of the
system are
\begin{equation*}
P^{(0)}\B{\chi}{i}{k}{j}{x_2,\infty}{x_1}=
 \begin{cases}
 \frac{1}{3}, & \chi = B, i = 0, j=0, k>0 \\
 0, & \text{otherwise}
  \end{cases} 
 \end{equation*}
 and $P_{1a}^{x_1}(0) = 1, P_{1b}^{x_1}(0) = 0, P_{9a}^{x_1}(0) = 1, P_{9b}^{x_1}(0) = 0$.

\begin{figure}[hp] 
   \centering
   \includegraphics[width=\textwidth]{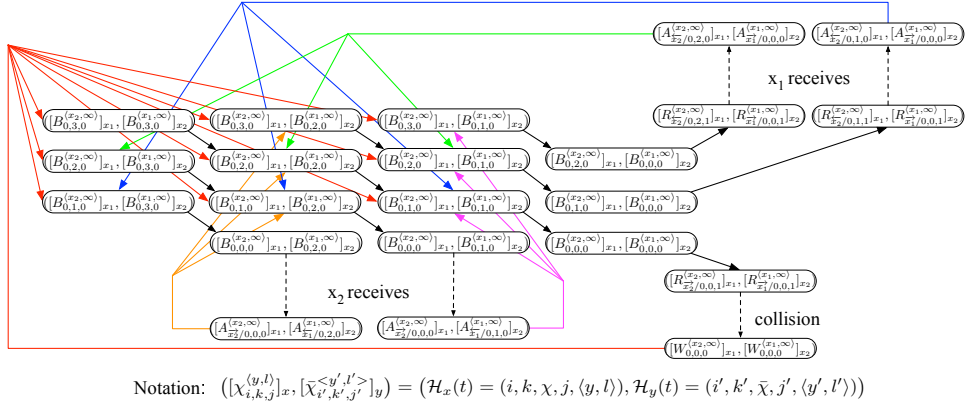} 
   \caption{2-node joint state model ($m = 0$)}
   \label{Ex1: Joint}
\end{figure}

\begin{figure}[hp] 
   \centering
   \includegraphics[width=\textwidth]{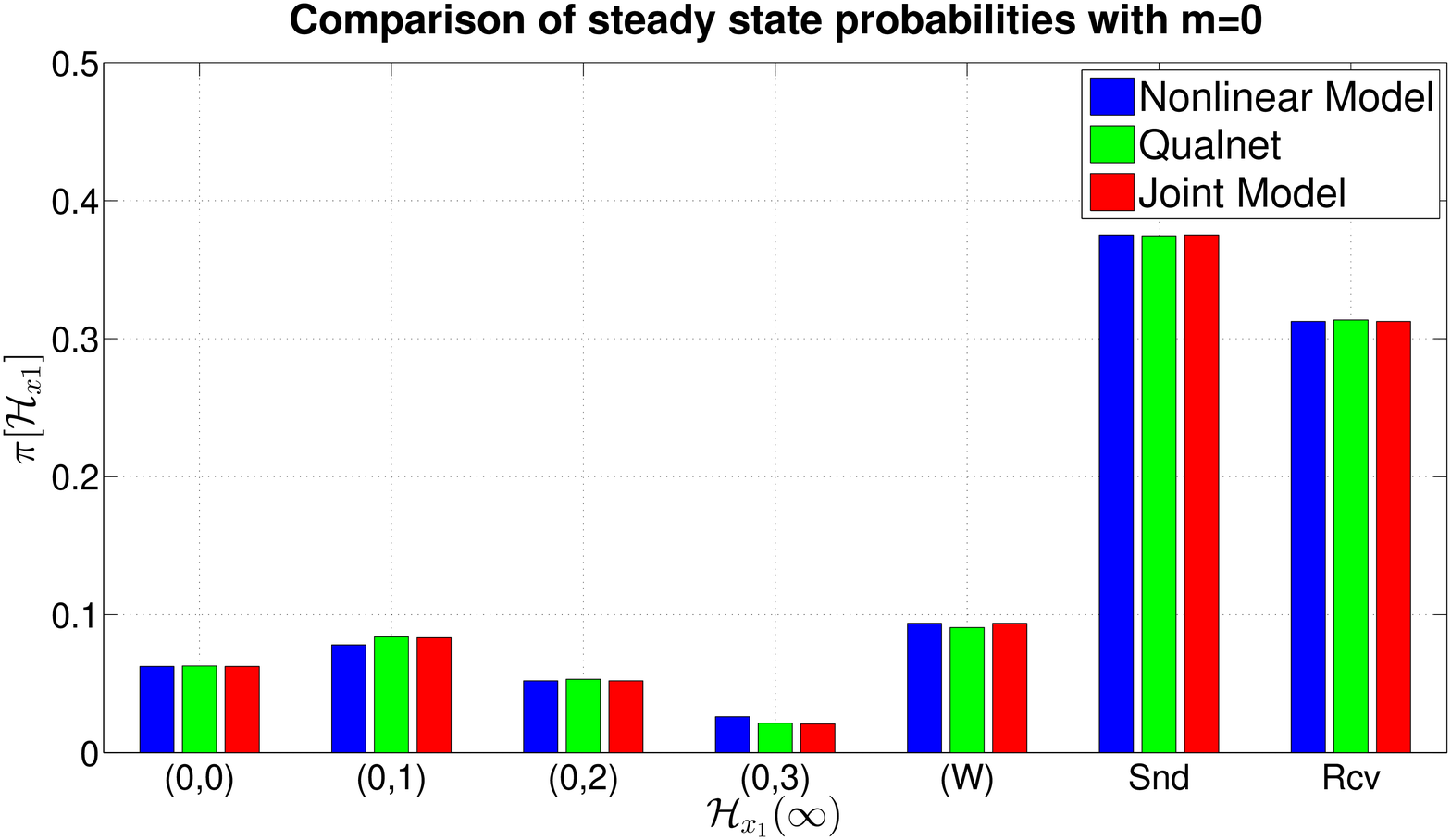} 
   \includegraphics[width=\textwidth]{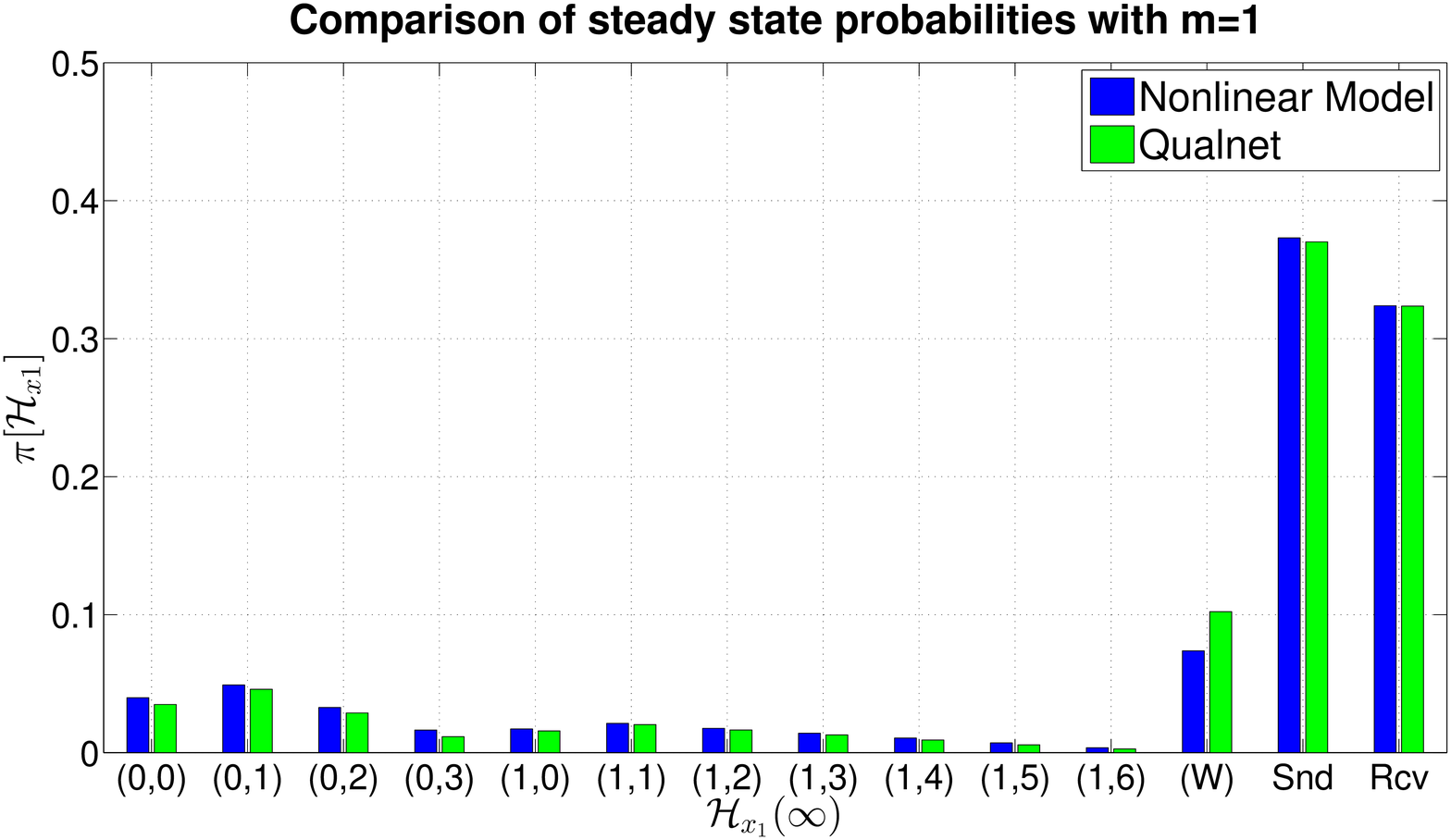} 
   \includegraphics[width=\textwidth]{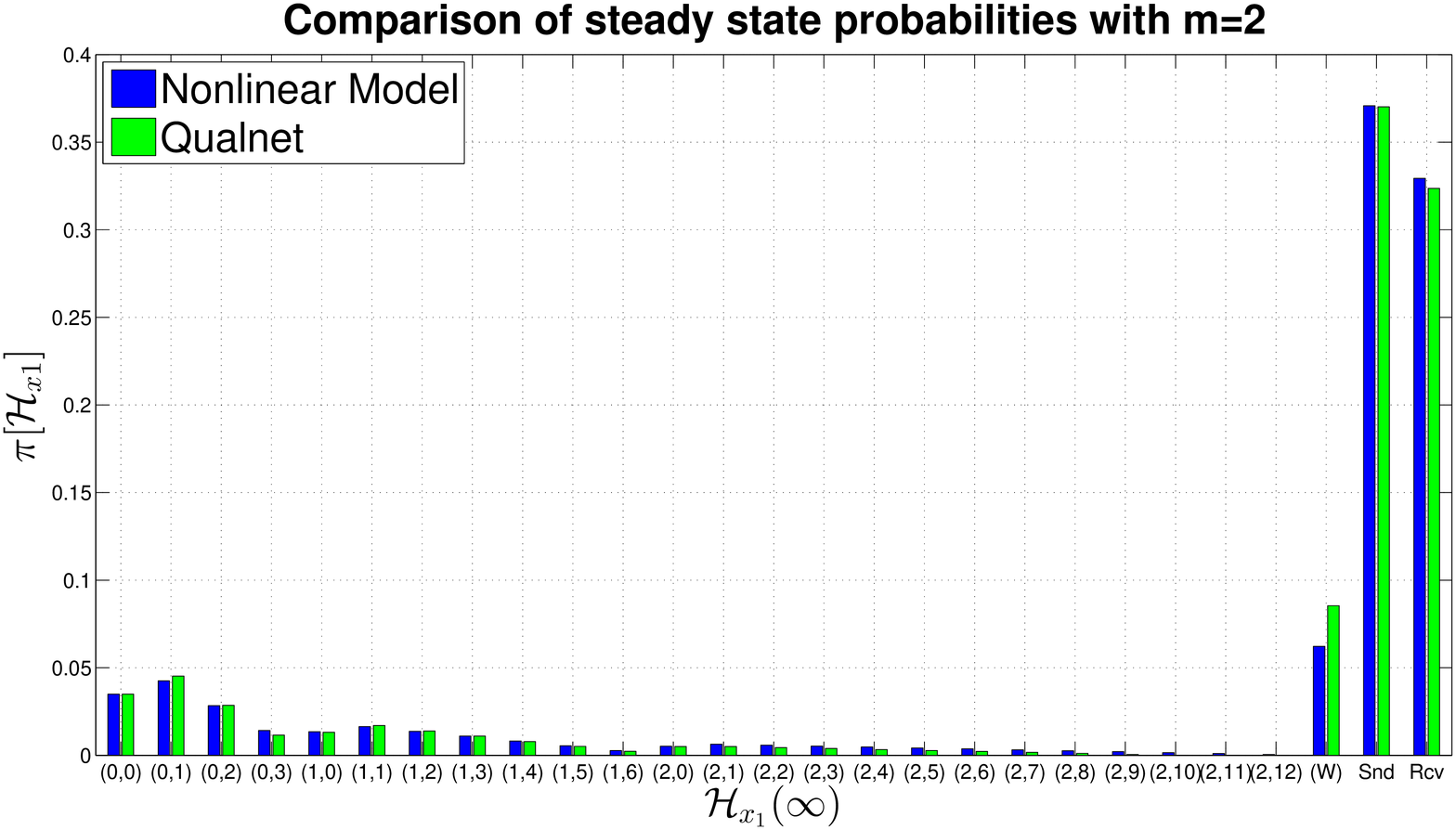} 
   
   \caption{Comparison at $x_1$: the tuples represent (back-off stage, back-off
counter); 'Snt' combines states of sending RTS/ receiving CTS/ sending DATA;
'Rcv' combines states of receiving RTS/ sending CTS/ receiving DATA
   \label{Ex1: 0-1-2 retry}}
\end{figure}
Figure \ref{Ex1: Joint} illustrates the dynamic among all the possible joint
states of $x_1$ and $x_2$ when no RTS retry is allowed ($m=0$). In the first
plot in Figure \ref{Ex1: 0-1-2 retry}, we compare the steady state distributions
attained separately by solving the non-linear system using the product
approximation (blue bars), exploring the joint state diagram (red bars) and
implementing QualNet simulation (green bars). Overall, the joint state diagram
accurately catch the nodes' behaviors and interactions under 802.11 DCF, and
more importantly, so does the nonlinear system representing our model. The
following two graphs in \ref{Ex1: 0-1-2 retry} show the results when we allow
RTS retransmits 1 or 2 times. The joint state model between $x_1$ and $x_2$
contains highly irregular structure and therefore is difficult to solve
directly. Nevertheless, our non-linear model closely reproduces the network
behavior captured by the QualNet simulation.

\subsection{A triangle network} 
As an interesting extension of the simple two-node system, we now consider an
equilateral triangle network where three nodes $x_1$, $x_2$ and $x_3$ share the
medium. Again, we assume each node has infinite packets in the queue and
randomly chooses a receiver. 
Due to symmetry, only $x_1$ will be considered.

Since there are no hidden nodes, transmissions always succeed unless two or
three nodes start to send simultaneously. As a consequence, only 8 transition
probability functions at time $t_n$ are non-trivial, summarized by the following
table. Notice in this example, $x_1$ may detect a busy channel during back-off
if $x_2$ and $x_3$ send a RTS at the same moment, resulting in a collision at
$x_1$.
\begin{table}[!h]
\footnotesize
 \caption{Non-Trivial Transition Probabilities}\label{t: Tp - tri}
\renewcommand{\arraystretch}{1.4}
\centering
\begin{tabular} {|  c  || p{5.5cm} | c || p{4.5cm} |}   
  \hline
  $\circled{$1_a$}$ & $x_1$ detects a quiet channel while back-off&$\circled{$8_a$}$&  $x_1$ detects the channel is clear \\ \hline
  $\circled{$1_b$}$&  $x_1$ detects a RTS while back-off &$\circled{$8_b$}$&  $x_1$ detects the channel is still busy  \\ \hline
  $\circled{$1_c$}$&  $x_1$ overhears a RTS while back-off  & $\circled{$9_a$}$ & RTS sent from $x_1$ succeeds\\ \hline
  $\circled{$1_e$}$&  $x_1$ detects an busy channel & $\circled{$9_b$}$ & RTS sent from $x_1$ fails\\ \hline  
\end{tabular}
\end{table}
By applying product approximation, we can evaluate the above probabilities in terms of the marginal densities of $x_2$ and $x_3$ as the following. 
\begin{align} \label{TP1a-3nodes}
 P_{1a}^{x_1}(t_n)&=\prod_{x_\alpha\in \{x_2,x_3\}}\TP{x_\alpha}{x_\alpha}{\sum_{\chi=B, k\neq 0}+\sum_{\chi\in\{W,U\}}}{\sum_{\chi\in\{B,W,U\}}}\\
 P_{1b}^{x_1x_2}(t_n) &= \TP{x_2}{x_2}{\sum_{\chi = B, k = 0, y=x_1}}{\sum_{\chi\in\{B,W,U\}}}\TP{x_3}{x_3}{\sum_{\chi=B, k\neq 0}+\sum_{\chi\in\{W,U\}}}{\sum_{\chi\in\{B,W,U\}}}\\
 P_{1b}^{x_1x_3}(t_n) &= \TP{x_3}{x_3}{\sum_{\chi = B, k = 0, y=x_1}}{\sum_{\chi\in\{B,W,U\}}}\TP{x_2}{x_2}{\sum_{\chi=B, k\neq 0}+\sum_{\chi\in\{W,U\}}}{\sum_{\chi\in\{B,W,U\}}}\\
 P_{1c}^{x_1x_2}(t_n) &= \TP{x_2}{x_2}{\sum_{\chi = B, k = 0, y=x_3}}{\sum_{\chi\in\{B,W,U\}}}\TP{x_3}{x_3}{\sum_{\chi=B, k\neq 0}+\sum_{\chi\in\{W,U\}}}{\sum_{\chi\in\{B,W,U\}}}\\
 P_{1c}^{x_1x_3}(t_n) &= \TP{x_3}{x_3}{\sum_{\chi = B, k = 0, y=x_2}}{\sum_{\chi\in\{B,W,U\}}}\TP{x_2}{x_2}{\sum_{\chi=B, k\neq 0}+\sum_{\chi\in\{W,U\}}}{\sum_{\chi\in\{B,W,U\}}}\\
 P_{1e}^{x_1}(t_n)&=1-P_{1a}^{x_1}(t_n)-P_{1b}^{x_1x_2}(t_n)-P_{1b}^{x_1x_3}(t_n)-P_{1c}^{x_1x_2}(t_n)-P_{1c}^{x_1x_3}(t_n)\\
 P_{8a}^{x_1}(t_n)&=\TP{x_2 \times x_3}{x_2\times x_3}{\sum_{\Omega_{8a}(\B{U}{i}{k}{0}{y,\infty}{x_1}; x_2, x_3)}}{\sum_{\Omega(\B{U}{i}{k}{0}{y,\infty}{x_1}; x_2, x_3)}}\\
 P_{8b}^{x_1}(t_n)&=1-P_{8a}^{x_1}(t_n)\\
 P_{9a}^{x_1x_2}(t_n)&=\TP{x_2}{x_1}{\sum_{\chi=R_{\rcv{x_1}},j=0}}{\sum_{\chi=R_{\snt{x_2}},j=0}}\\
P_{9b}^{x_1x_2}(t_n)&=1-P_{9a}^{x_1x_2}(t_n)\\
P_{9a}^{x_1x_3}(t_n)&=\TP{x_3}{x_1}{\sum_{\chi=R_{\rcv{x_1}},j=0}}{\sum_{\chi=R_{\snt{x_3}},j=0}}\\ 
P_{9b}^{x_1x_3}(t_n)&=1-P_{9a}^{x_1x_3}(t_n) \label{TP9b-3nodes}
 \end{align}
where $\Omega(\B{U}{i}{k}{0}{y,\infty}{x_1}; x_2, x_3)$ represents
\begin{multline*}
\{\mathcal{H}_{x_2}(t_{n})|\chi_{x_2}\in\{(R/A)_{\snt{x_1}/\snt{x_3}},C_{\snt{x_3}}\}\}\times\{\mathcal{H}_{x_3}(t_{n})|\chi_{x_3}\notin\{R_{\rcv{x_1}/\ovh{x_1}}, C_{\snt{x_1}/\rcv{x_1}/\ovh{x_1}}, A_{\rcv{x_1}}\}\}\\
\quad\cup\{\mathcal{H}_{x_3}(t_{n})|\chi_{x_3}\in\{(R/A)_{\snt{x_1}/\snt{x_2}},C_{\snt{x_2}}\}\}\times\{\mathcal{H}_{x_2}(t_{n})|\chi_{x_2}\notin\{R_{\rcv{x_1}/\ovh{x_1}}, C_{\snt{x_1}/\rcv{x_1}/\ovh{x_1}}, A_{\rcv{x_1}}\}\}
\end{multline*}
and $\Omega_{8a}(\B{U}{i}{k}{0}{y,\infty}{x_1}; x_2, x_3)$ stands for
\begin{multline*}
\{\mathcal{H}_{x_2}(t_{n})\in\Omega(\B{U}{i}{k}{0}{y,\infty}{x_1}; x_2)| j_{x_2}=0\}\times\{\mathcal{H}_{x_3}(t_{n})\in \Omega(\B{U}{i}{k}{0}{y,\infty}{x_1}; x_2; x_3)|\\
(\chi_{x_3},j_{x_3})\notin\{((R/C/A)_{\snt{x_1}/\snt{x_2}},j)\}, j\ne 0, (\chi_{x_3},k_{x_3})\ne (B,0)\}\\
\cup\{\mathcal{H}_{x_3}(t_{n})\in\Omega(\B{U}{i}{k}{0}{y,\infty}{x_1}; x_3)| j_{x_3}=0\}\times\{\mathcal{H}_{x_2}(t_{n})\in \Omega(\B{U}{i}{k}{0}{y,\infty}{x_1}; x_3; x_2)|\\
(\chi_{x_2},j_{x_2})\notin\{((R/C/A)_{\snt{x_1}/\snt{x_3}},j)\}, j\ne 0, (\chi_{x_2},k_{x_2})\ne (B,0)\}
\end{multline*}
The stationary distribution of back-off states as $t_n\to \infty$ satisfy:
\begin{align}
\pi\B{B}{i}{0}{0}{y,\infty}{x_1}&=p^{x_1}_{1a} \pi\B{B}{i}{1}{0}{y,\infty}{x_1}\\
\pi\B{B}{i}{k}{0}{y,\infty}{x_1}&=
\begin{cases}
 	p^{x_1}_{1a} \pi\B{B}{i}{k+1}{0}{y,\infty}{x_1}+(1-p^{x_1}_{1a}) \pi\B{B}{i}{k}{0}{y,\infty}{x}, & 0<k<2^i*3\\
	(1-p^{x_1}_{1a})\pi\B{B}{i}{2^iw}{0}{y,\infty}{x}, & k=2^i*3
\end{cases} \notag\\
 & \quad+\begin{cases}
	\frac{P_{x_1y}}{2^i*3} (p_{9b}^{x_1x_2}\pi\B{B}{i-1}{0}{0}{x_2,\infty}{x_1}+p_{9b}^{x_1x_3}\pi\B{B}{i-1}{0}{0}{x_3,\infty}{x_1}), & i>0, k\ne 0\\
	\frac{P_{x_1y}}{3}(p^{x_1x_2}_{9a}\sum_{i=0}^{m-1} \pi\B{B}{i}{0}{0}{x_2,\infty}{x_1}+\pi\B{B}{m}{0}{0}{x_2,\infty}{x_1}&\\
	\qquad +p^{x_1x_3}_{9a}\sum_{i=0}^{m-1} \pi\B{B}{i}{0}{0}{x_3,\infty}{x_1}+\pi\B{B}{m}{0}{0}{x_3,\infty}{x_1}), & i=0, k\ne 0\\
	\end{cases}\label{steadyequ: Bs_ex2}
\end{align} 
for any $y\in\{x_2,x_3\}$. Note we assume at the beginning of each new DATA
session, the sender chooses it's receiver randomly, thus
$P_{x_1y}=\frac{1}{|N_{x_1}|}=\frac{1}{2}$. In general, this will be set as a
parameter that is determined by the routing algorithm or experimental settings.

For the remaining part of the distribution, one can conclude that
\begin{align}
\pi\B{A}{\rcv{x_2}/i}{k}{0}{y,\infty}{x_1}&=\cdots=\pi\B{A}{\rcv{x_2}/i}{k}{5}{y,\infty}{x_1}= \pi\B{C}{\snt{x_2}/ i}{k}{0}{y,\infty}{x_1}=\cdots= \pi\B{R}{\rcv{x_2}/i}{k}{1}{y,\infty}{x_1}= p^{x_1x_2}_{1b} \pi\B{B}{i}{k}{0}{y,\infty}{x_1}\notag\\
\pi\B{A}{\rcv{x_3}/i}{k}{0}{y,\infty}{x_1}&=\cdots=\pi\B{A}{\rcv{x_3}/i}{k}{5}{y,\infty}{x_1}= \pi\B{C}{\snt{x_3}/ i}{k}{0}{y,\infty}{x_1}=\cdots= \pi\B{R}{\rcv{x_3}/i}{k}{1}{y,\infty}{x_1}= p^{x_1x_3}_{1b} \pi\B{B}{i}{k}{0}{y,\infty}{x_1}\\
\pi\B{D}{{x_2}/i}{k}{0}{y,\infty}{x_1}&=\cdots=\pi\B{D}{{x_2}/i}{k}{7}{y,\infty}{x_1}= \pi\B{R}{\ovh{x_2}/ i}{k}{0}{y,\infty}{x_1}= \pi\B{R}{\ovh{x_2}/i}{k}{1}{y,\infty}{x_1}= p^{x_1x_2}_{1c} \pi\B{B}{i}{k}{0}{y,\infty}{x_1}\notag\\
\pi\B{D}{{x_3}/i}{k}{0}{y,\infty}{x_1}&=\cdots=\pi\B{D}{{x_3}/i}{k}{7}{y,\infty}{x_1}= \pi\B{R}{\ovh{x_3}/ i}{k}{0}{y,\infty}{x_1}=\pi\B{R}{\ovh{x_3}/i}{k}{1}{y,\infty}{x_1}= p^{x_1x_3}_{1c} \pi\B{B}{i}{k}{0}{y,\infty}{x_1} \quad \\
\pi\B{U}{i}{k}{0}{y,\infty}{x_1}&=\frac{p^{x_1}_{1e}}{p^{x_1}_{8a}}\pi\B{B}{i}{k}{0}{y,\infty}{x_1}\\
\pi\B{A}{\snt{x_2}/0}{0}{0}{y,\infty}{x_1}&= \cdots =\sum_{i=0}^{m}\pi\B{C}{\rcv{x_2}/i}{0}{1}{y,\infty}{x_1}=\sum_{i=0}^{m} p^{x_1x_2}_{9a}\pi\B{R}{\snt{x_2}/i}{0}{0}{y,\infty}{x_1}=\cdots =\sum_{i=0}^{m}p^{x_1x_2}_{9a} \pi\B{B}{i}{0}{0}{y,\infty}{x_1}\notag\\
\pi\B{A}{\snt{x_3}/0}{0}{0}{y,\infty}{x_1}&= \cdots =\sum_{i=0}^{m}\pi\B{C}{\rcv{x_3}/i}{0}{1}{y,\infty}{x_1}=\sum_{i=0}^{m} p^{x_1x_3}_{9a}\pi\B{R}{\snt{x_3}/i}{0}{0}{y,\infty}{x_1}=\cdots =\sum_{i=0}^{m}p^{x_1x_3}_{9a} \pi\B{B}{i}{0}{0}{y,\infty}{x_1}\\
\pi\B{W}{i}{0}{0}{y,\infty}{x_1}&=\cdots=\pi\B{W}{i}{0}{2}{y,\infty}{x_1}=p^{x_1y}_{9b}\pi\B{R}{\snt{y}/i}{0}{0}{y,\infty}{x_1}=\cdots=p^{x_1y}_{9b}\pi\B{B}{i}{0}{0}{y,\infty}{x_1}\label{steadyequ: Os_ex2}
\end{align} 
Given the transition probability functions from \eqref{TP1a-3nodes} to \eqref{TP9b-3nodes}, together with the condition of symmetric topology: 
 \begin{equation}
 \pi\B{\chi}{i}{k}{j}{y,\infty}{x_1}=\pi\B{\chi}{i}{k}{j}{y,\infty}{x_2}=\pi\B{\chi}{i}{k}{j}{y,\infty}{x_3}
  \end{equation}
 for any $i,k,\chi,j,y$, and the normalization conditions
 \begin{equation}
 \sum_{i,k,\chi,j,y}\pi\B{\chi}{i}{k}{j}{y,\infty}{x_1}=\sum_{i,k,\chi,j,y}\pi\B{\chi}{i}{k}{j}{y,\infty}{x_2}=\sum_{i,k,\chi,j,y}\pi\B{\chi}{i}{k}{j}{y,\infty}{x_3}=1
 \end{equation}
we form a non-linear system which can be solved again by Matlab with initial conditions:
\begin{equation*}
P^{(0)}\B{\chi}{i}{k}{j}{y,\infty}{x_1}=
 \begin{cases}
 \frac{1}{6}, & \chi = B, i = 0, j=0, k>0 \\
 0, & \text{otherwise}
  \end{cases} 
 \end{equation*}
 and $P_{1a}^{x_1}(0) = P_{8a}^{x_1}(0) =  P_{9a}^{x_1x_2}(0) = P_{9a}^{x_1x_3}(0) = 1$. The remaining significant transition probabilities are initialized as 0.

\begin{figure}[hp] 
   \centering
   \includegraphics[width=\textwidth]{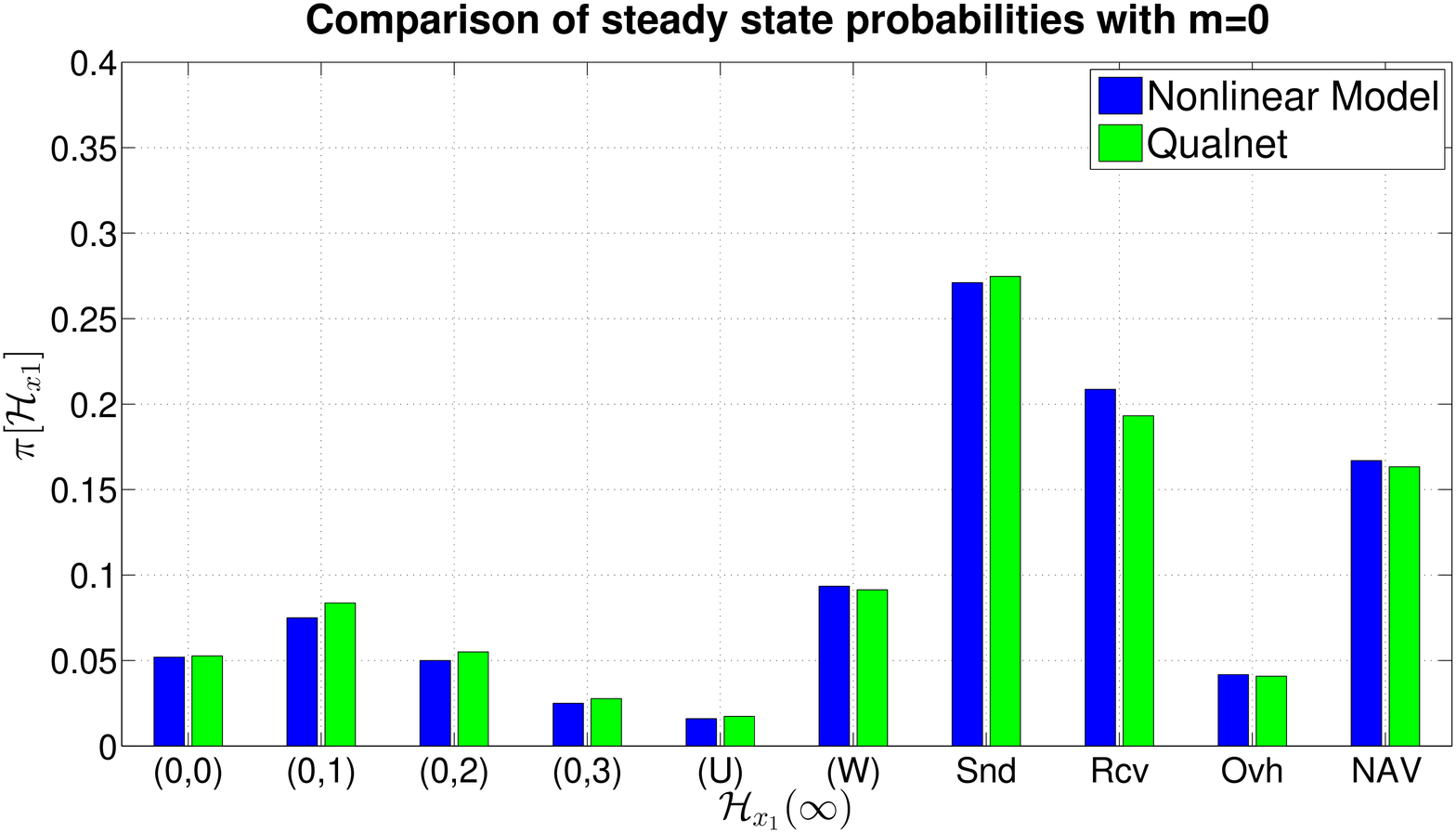} 
   \includegraphics[width=\textwidth]{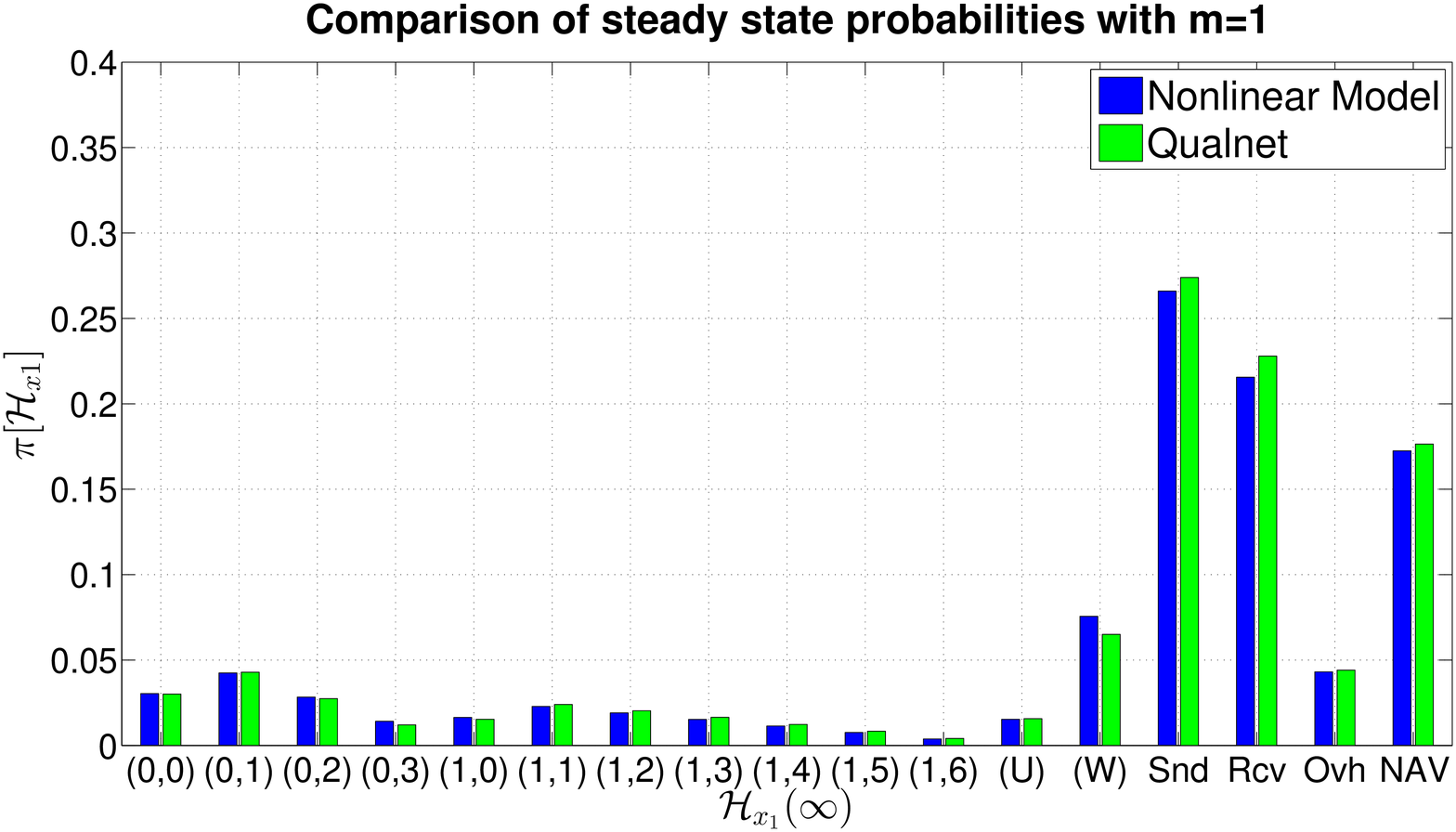} 
   \includegraphics[width=\textwidth]{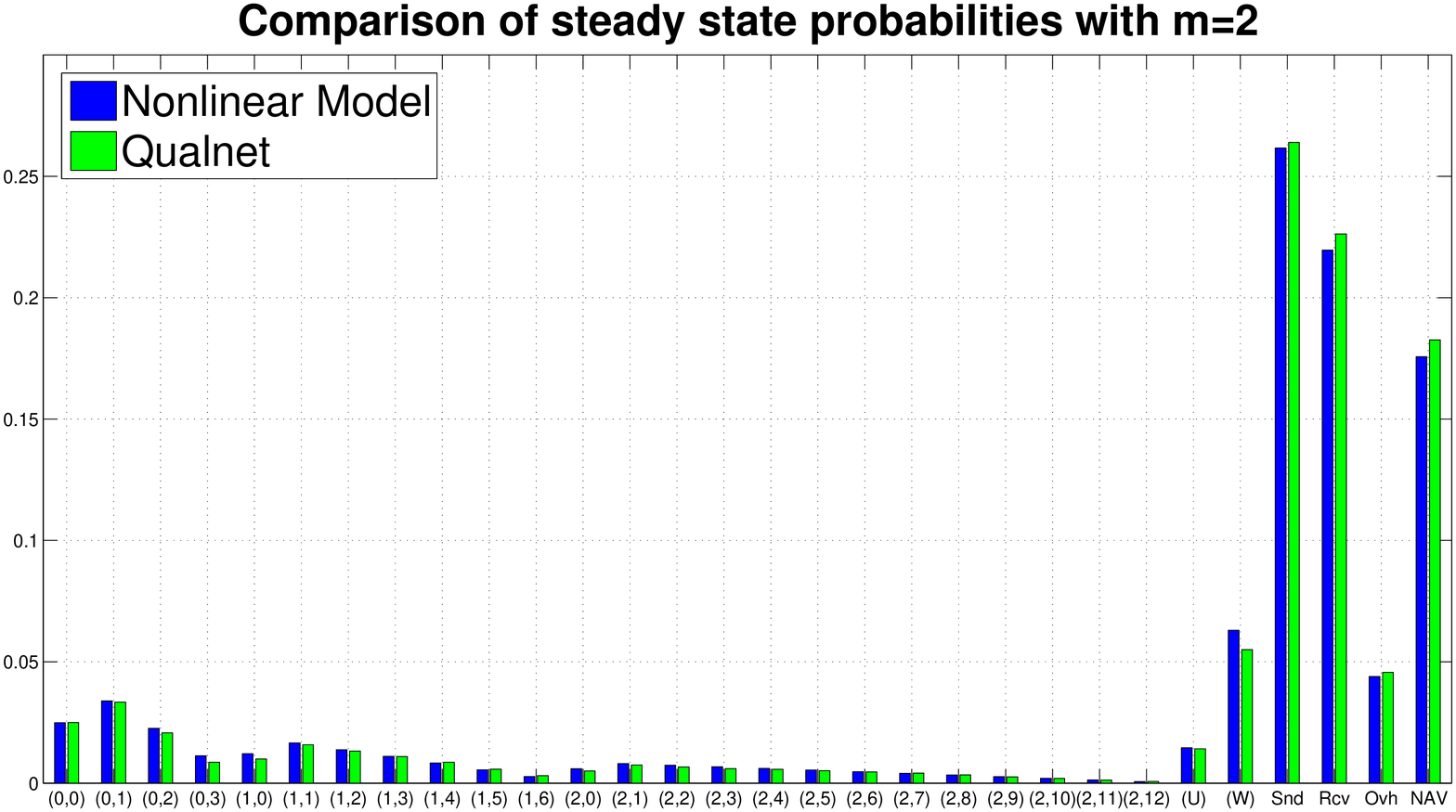}
   \caption{Comparison at $x_1$: the tuples represent (back-off stage, back-off
counter); 'Snt' combines states of sending RTS/ receiving CTS/ sending DATA;
'Rcv' combines states of receiving RTS/ sending CTS/ receiving DATA; 'Ovh'
denotes overhearing RTS\label{Ex2: 0-1-2 retry}}
\end{figure}

The plots in Figure \ref{Ex2: 0-1-2 retry}  compare the steady
state distributions with analytical results from the above system and QualNet
simulations where the RTS retransmission limit ($m$) is set to be $0, 1, 2$
respectively. 

\subsection{2 Senders and 1 receiver}
Finally we examine the following topology where three nodes, $x_1$, $x_2$ and
$x_3$, are presented in the network. Both $x_1$ and $x_3$ have infinite data
packets destined to node $x_2$ in their queues.  This case includes hidden
terminals because node $x_3$ is not within range of $x_1$ and vice versa. Node
$x_2$ has no queue and absorbs data packet from $x_1$ and $x_3$. 
Due to symmetry, only $x_1$ and $x_2$ will be considered. 

Unlike the previous example, $x_1$ and $x_3$ are hidden to each other thus the
receiving procedure of RTS at $x_2$ may be interrupted by collision. On the
other hand the receiving of CTS packets at $x_1$ always succeeds since $x_1$ has
no other neighbors (beside $x_2$) to interfere. Notice this fact indicates all
the subsequent data packets (in this case, from $x_3$ to $x_2$) will be
protected by NAV hence probability $\circled{$6_a$}\equiv 1$.

\begin{table}[!h]
\footnotesize
\renewcommand{\arraystretch}{1.4}
 \caption{Non-Trivial Transition Probabilities for $x_1$}\label{t4}
\centering
\begin{tabular} {|  c  || p{0.4 \columnwidth}  |}   
  \hline
  $\circled{$1_a$}$ & $x_1$ detects a quiet channel while back-off\\ \hline
  $\circled{$1_d$}$&  $x_1$ overhears a CTS while back-off \\ \hline
  $\circled{$9_a$}$ & RTS sent from $x_1$ succeeds\\ \hline
  $\circled{$9_b$}$ & RTS sent from $x_1$ fails \\ \hline
\end{tabular}

 \caption{Non-Trivial Transition Probabilities for $x_2$}\label{t5}
\begin{tabular} {|  c  || p{.45 \columnwidth}  |}   
  \hline
  $\circled{$1_a$}$ & $x_2$ detects a quiet channel while idle\\ \hline
  $\circled{$1_b$}$&  $x_2$ detects a RTS while idle \\ \hline  
  $\circled{$1_e$}$&  $x_2$ detects a busy channel while idle \\ \hline
  $\circled{$2_a$}$ & $x_2$ receives RTS correctly\\ \hline
  $\circled{$2_b$}$ & $x_2$ detects a collision while receiving RTS \\ \hline
  $\circled{$8_a$}$&  $x_2$ detects the channel is clear \\ \hline
  $\circled{$8_b$}$&  $x_2$ detects the channel is still busy \\ \hline
\end{tabular}
  
\end{table}
The only possible non-trivial transition probability functions at time $t_n$ are summarized by table \ref{t4} and \ref{t5}. For node $x_1$ we have
\begin{align} \label{TP1a-3nodesH}
  P_{1a}^{x_1}(t_n)&= \TP{x_2}{x_2}{\sum_{\chi=R_{\rcv{x_3}}, j\ne 0}+\sum_{\chi\in\{I,U\}}}{\sum_{\chi\in\{I, R_{\rcv{x_3}},U\}}}\\
  P_{1d}^{x_1x_2}(t_n)&=\TP{x_2}{x_2}{\sum_{\chi=R_{\rcv{x_3}}, j = 0}}{\sum_{\chi\in\{I, R_{\rcv{x_3}},U\}}}\\
  P_{9a}^{x_1x_2}(t_n)&=\TP{x_2}{x_1}{\sum_{\chi=R_{\rcv{x_1}},j=0}}{\sum_{\chi=R_{\snt{x_2}},j=0}}\\
  P_{9b}^{x_1x_2}(t_n)&=1-P_{9a}^{x_1x_2}(t_n)
 \end{align}
For node $x_2$, we have
\begin{align} \label{TP1a-3nodesH2}
  P_{1a}^{x_2}(t_n) &= \prod_{x_\alpha\in \{x_1,x_3\}}\TP{x_\alpha}{x_\alpha}{\sum_{\chi=B, k\neq 0}+\sum_{\chi = W}}{\sum_{\chi\in\{B,W\}}}\\
  P_{1b}^{x_2x_1}(t_n) &= \TP{x_1}{x_1}{\sum_{\chi = B, k = 0, y=x_2}}{\sum_{\chi\in\{B,W\}}}\TP{x_3}{x_3}{\sum_{\chi=B, k\neq 0}+\sum_{\chi = W}}{\sum_{\chi\in\{B,W\}}}\\
 P_{1b}^{x_2x_3}(t_n) &= \TP{x_3}{x_3}{\sum_{\chi = B, k = 0, y=x_2}}{\sum_{\chi\in\{B,W\}}}\TP{x_1}{x_1}{\sum_{\chi=B, k\neq 0}+\sum_{\chi = W}}{\sum_{\chi\in\{B,W\}}}\\
  P_{1e}^{x_2}(t_n)&=1- P_{1a}^{x_2}(t_n) - P_{1b}^{x_2x_1}(t_n) - P_{1b}^{x_2x_3}(t_n) \\
  P_{2a}^{x_2x_1}(t_n) &=\TP{x_3}{x_3}{\sum_{\chi=B, k\ne 0}+\sum_{\chi=W}}{\sum_{\chi\in\{B,W\}}}\\
  P_{2b}^{x_2x_1}(t_n)&=1-P_{2a}^{x_2x_1}(t_n)\\
   P_{2a}^{x_2x_3}(t_n) &=\TP{x_1}{x_1}{\sum_{\chi=B, k\ne 0}+\sum_{\chi=W}}{\sum_{\chi\in\{B,W\}}}\\
  P_{2b}^{x_2x_3}(t_n)&=1-P_{2a}^{x_2x_3}(t_n)\\
  P_{8a}^{x_2}(t_n)&=\TP{x_1\times x_3}{x_1\times x_3}{\sum_{\Omega_{8a}(\B{U}{0}{0}{0}{\emptyset,0}{x_2}; x_1, x_3)}}{\sum_{\Omega(\B{U}{0}{0}{0}{\emptyset,0}{x_2}; x_1,x_3)} }\\
  P_{8b}^{x_2}(t_n)&=1-P_{8a}^{x_2}(t_n)
 \end{align}
where $\Omega(\B{U}{0}{0}{0}{\emptyset,0}{x_2}; x_1,x_3)$ represents
\begin{multline*}
\{\mathcal{H}_{x_1}(t_{n})|\chi_{x_1}\in\{R_{\snt{x_2}}, A_{\snt{x_2}}\}\}\times\{\mathcal{H}_{x_3}(t_{n})|\chi_{x_3}\notin\{C_{\rcv{x_2}},C_{\ovh{x_2}}\}\}\\
\quad\cup\{\mathcal{H}_{x_3}(t_{n})|\chi_{x_3}\in\{R_{\snt{x_2}}, A_{\snt{x_2}}\}\}\times\{\mathcal{H}_{x_1}(t_{n})|\chi_{x_1}\notin\{C_{\rcv{x_2}},C_{\ovh{x_2}}\}\}
\end{multline*}
and $\Omega_{8a}(\B{U}{0}{0}{0}{\emptyset,0}{x_2}; x_1, x_3)$ stands for
\begin{multline*}
\{\mathcal{H}_{x_1}(t_{n})\in\Omega(\B{U}{0}{0}{0}{\emptyset,0}{x_2}; x_1)| j_{x_1}=0\}\times\{\mathcal{H}_{x_3}(t_{n})\in \Omega(\B{U}{0}{0}{0}{\emptyset,0}{x_2}; x_1; x_3)|\\(\chi_{x_3},j_{x_3})\notin\{(R_{\snt{x_2}},1),(A_{\snt{x_2}},j)\}, j\ne 0, (\chi_{x_3},k_{x_3})\ne (B,0)\}\\
\cup\{\mathcal{H}_{x_3}(t_{n})\in\Omega(\B{U}{0}{0}{0}{\emptyset,0}{x_2}; x_3)| j_{x_3}=0\}\times\{\mathcal{H}_{x_1}(t_{n})\in \Omega(\B{U}{0}{0}{0}{\emptyset,0}{x_2}; x_3; x_1)|\\(\chi_{x_1},j_{x_1})\notin\{(R_{\snt{x_2}},1),(A_{\snt{x_2}},j)\},j \ne 0, (\chi_{x_1},k_{x_1})\ne (B,0)\}
\end{multline*}
The equilibrium equations for $x_1$ as $t_n\to \infty$ are 
\begin{align} \label{steadyequ: Bs_ex3-x1}
\pi\B{B}{i}{0}{0}{x_2,\infty}{x_1}&=p^{x_1}_{1a} \pi\B{B}{i}{1}{0}{x_2,\infty}{x_1}\\
\pi\B{B}{i}{k}{0}{x_2,\infty}{x_1}&=
\begin{cases}
 	p^{x_1}_{1a} \pi\B{B}{i}{k+1}{0}{x_2,\infty}{x_1}+(1-p^{x_1}_{1a}) \pi\B{B}{i}{k}{0}{x_2,\infty}{x_1}, & 0<k<2^i*3\\
	(1-p^{x_1}_{1a})\pi\B{B}{i}{2^iw}{0}{x_2,\infty}{x_1}, & k=2^i*3
\end{cases} \notag\\
 & \quad+\begin{cases}
	\frac{1}{2^i*3} p_{9b}^{x_1x_2}\pi\B{B}{i-1}{0}{0}{x_2,\infty}{x_1}, & i>0, k\ne 0\\
	\frac{1}{3}(p^{x_1x_2}_{9a}\sum_{i=0}^{m-1} \pi\B{B}{i}{0}{0}{x_2,\infty}{x_1}+\pi\B{B}{m}{0}{0}{x_2,\infty}{x_1}), & i=0, k\ne 0\\
	\end{cases}\\
\pi\B{D}{{x_2}/i}{k}{0}{x_2,\infty}{x_1}&=\cdots=\pi\B{D}{{x_2}/i}{k}{7}{x_2,\infty}{x_1}= \pi\B{C}{\ovh{x_2}/ i}{k}{0}{x_2,\infty}{x_1}= \pi\B{C}{\ovh{x_2}/i}{k}{1}{x_2,\infty}{x_1}= p^{x_1x_2}_{1d} \pi\B{B}{i}{k}{0}{x_2,\infty}{x_1}\\
\pi\B{A}{\snt{x_2}/0}{0}{0}{x_2,\infty}{x_1}&= \cdots =\sum_{i=0}^{m}\pi\B{C}{\rcv{x_2}/i}{0}{1}{x_2,\infty}{x_1}=\sum_{i=0}^{m} p^{x_1x_2}_{9a}\pi\B{R}{\snt{x_2}/i}{0}{0}{x_2,\infty}{x_1}=\cdots =\sum_{i=0}^{m}p^{x_1x_2}_{9a} \pi\B{B}{i}{0}{0}{x_2,\infty}{x_1}\\
\pi\B{W}{i}{0}{0}{x_2,\infty}{x_1}&=\cdots=\pi\B{W}{i}{0}{2}{x_2,\infty}{x_1}=p^{x_1x_2}_{9b}\pi\B{R}{\snt{x_2}/i}{0}{0}{x_2,\infty}{x_1}=\cdots=p^{x_1x_2}_{9b}\pi\B{B}{i}{0}{0}{x_2,\infty}{x_1}
\end{align} 
The equilibrium equations for $x_2$ follows (\ref{steadyequ-I}), (\ref{steadyequ-U}), (\ref{steadyequ-A}):
\begin{align}
\pi\B{I}{0}{0}{0}{\emptyset,0}{x_2}&=\frac{1}{1-p^{x_2}_{1a}}( p^{x_2}_{8a}\pi\B{U}{0}{0}{0}{\emptyset, 0}{x_2}+\pi\B{A}{\rcv{x_1}/0}{0}{0}{\emptyset, 0}{x_2}+\pi\B{A}{\rcv{x_3}/0}{0}{0}{\emptyset, 0}{x_2})\\
\pi\B{A}{\rcv{x_1}/0}{0}{0}{\emptyset,0}{x_2}&=\cdots=\pi\B{R}{\rcv{x_1}/0}{0}{0}{\emptyset,0}{x_2}=p^{x_2x_1}_{2a} \pi\B{R}{\rcv{x_1}/0}{0}{1}{\emptyset,0}{x_2}= p^{x_2x_1}_{2a}p^{x_2x_1}_{1b} \pi\B{I}{0}{0}{0}{\emptyset,0}{x_2}\notag\\
\pi\B{A}{\rcv{x_3}/0}{0}{0}{\emptyset,0}{x_2}&=\cdots=\pi\B{R}{\rcv{x_3}/0}{0}{0}{\emptyset,0}{x_2}=p^{x_2x_3}_{2a} \pi\B{R}{\rcv{x_3}/0}{0}{1}{\emptyset,0}{x_2}= p^{x_2x_3}_{2a}p^{x_2x_3}_{1b} \pi\B{I}{0}{0}{0}{\emptyset,0}{x_2}\\
\pi\B{U}{i}{k}{0}{y,\infty}{x_1}&=\frac{1}{p^{x_2}_{8a}}(p^{x_2}_{1e}\pi\B{I}{0}{0}{0}{\emptyset,0}{x_2}+p^{x_2x_1}_{8b}\pi\B{R}{\rcv{x_1}/0}{0}{1}{\emptyset,0}{x_2}+p^{x_2x_3}_{8b}\pi\B{R}{\rcv{x_3}/0}{0}{1}{\emptyset,0}{x_2}) \label{steadyequ: Bs_ex3-x2}
\end{align} 
Finally, using the symmetric conditions 
 \begin{equation}
 \pi\B{\chi}{i}{k}{j}{x_2,\infty}{x_1}=\pi\B{\chi}{i}{k}{j}{x_2,\infty}{x_3}
  \end{equation}
 for any $i,k,\chi,j$, and the normalization conditions
 \begin{equation}\label{no}
 \sum_{i,k,\chi,j}\pi\B{\chi}{i}{k}{j}{x_2,\infty}{x_1}=\sum_{i,k,\chi,j}\pi\B{\chi}{i}{k}{j}{\emptyset,0}{x_2}=\sum_{i,k,\chi,j}\pi\B{\chi}{i}{k}{j}{x_2,\infty}{x_3}=1
 \end{equation}
 we form a non-linear system by combing equations (\ref{TP1a-3nodesH}) - (\ref{steadyequ: Bs_ex3-x2}). The initial conditions for $x_1$ are similar to the settings in section \ref{sec: 2-node}. For $x_2$, we have
 \begin{equation*}
P^{(0)}\B{\chi}{0}{0}{j}{\emptyset,0}{x_1}=
 \begin{cases}
 1, & \chi = I \\
 0, & \text{otherwise}
  \end{cases} 
 \end{equation*}
 and $P_{1a}^{x_2}(0) = P_{2a}^{x_2x_1}(0) =  P_{2a}^{x_2x_3}(0) = P_{8a}^{x_2}(0) = 1$ while the remaining non-trivial transition probabilities are initialized as 0.

\begin{figure}[hp] 
   \centering
   \includegraphics[width=\textwidth]{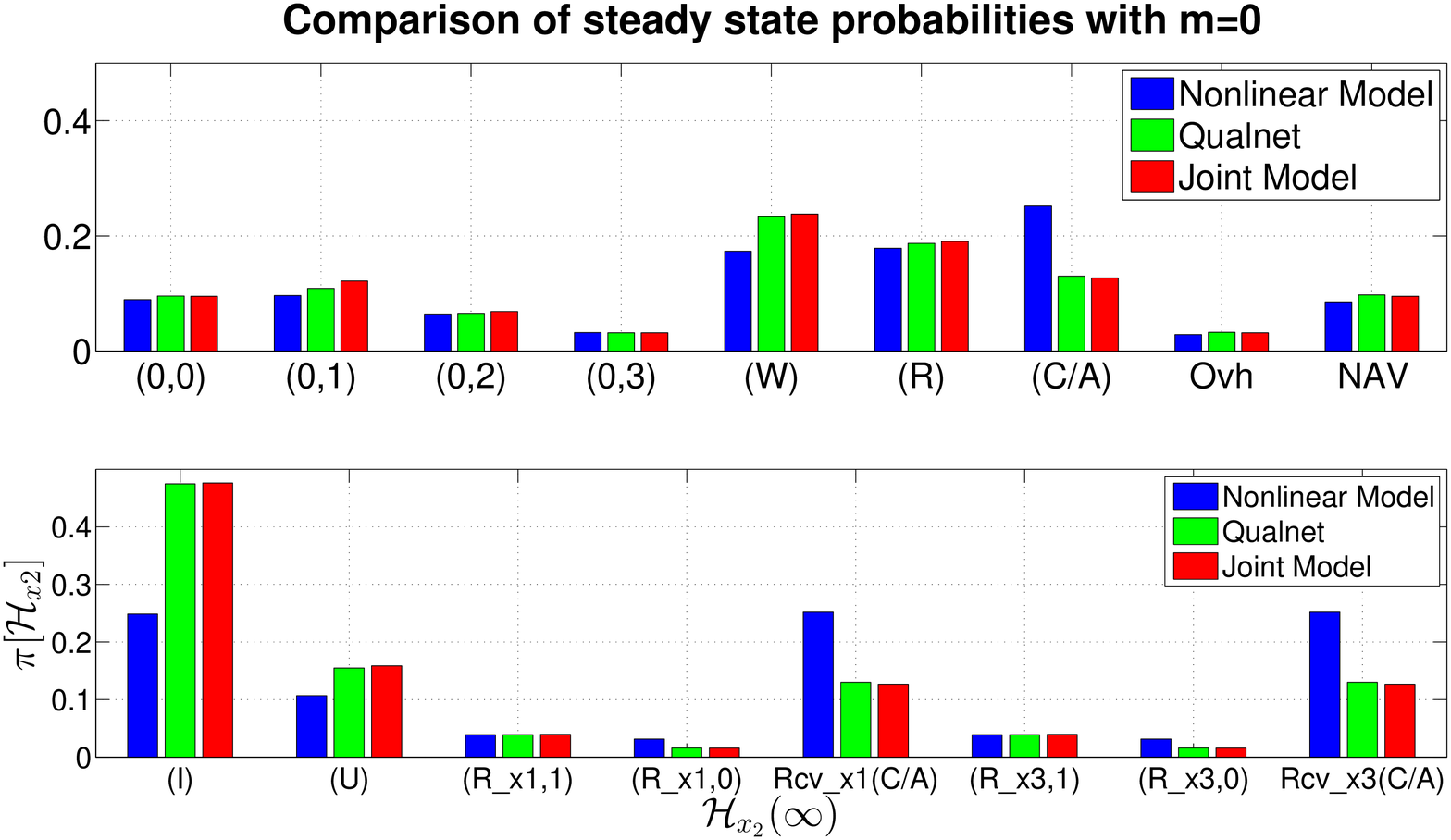} 
   \includegraphics[width=\textwidth]{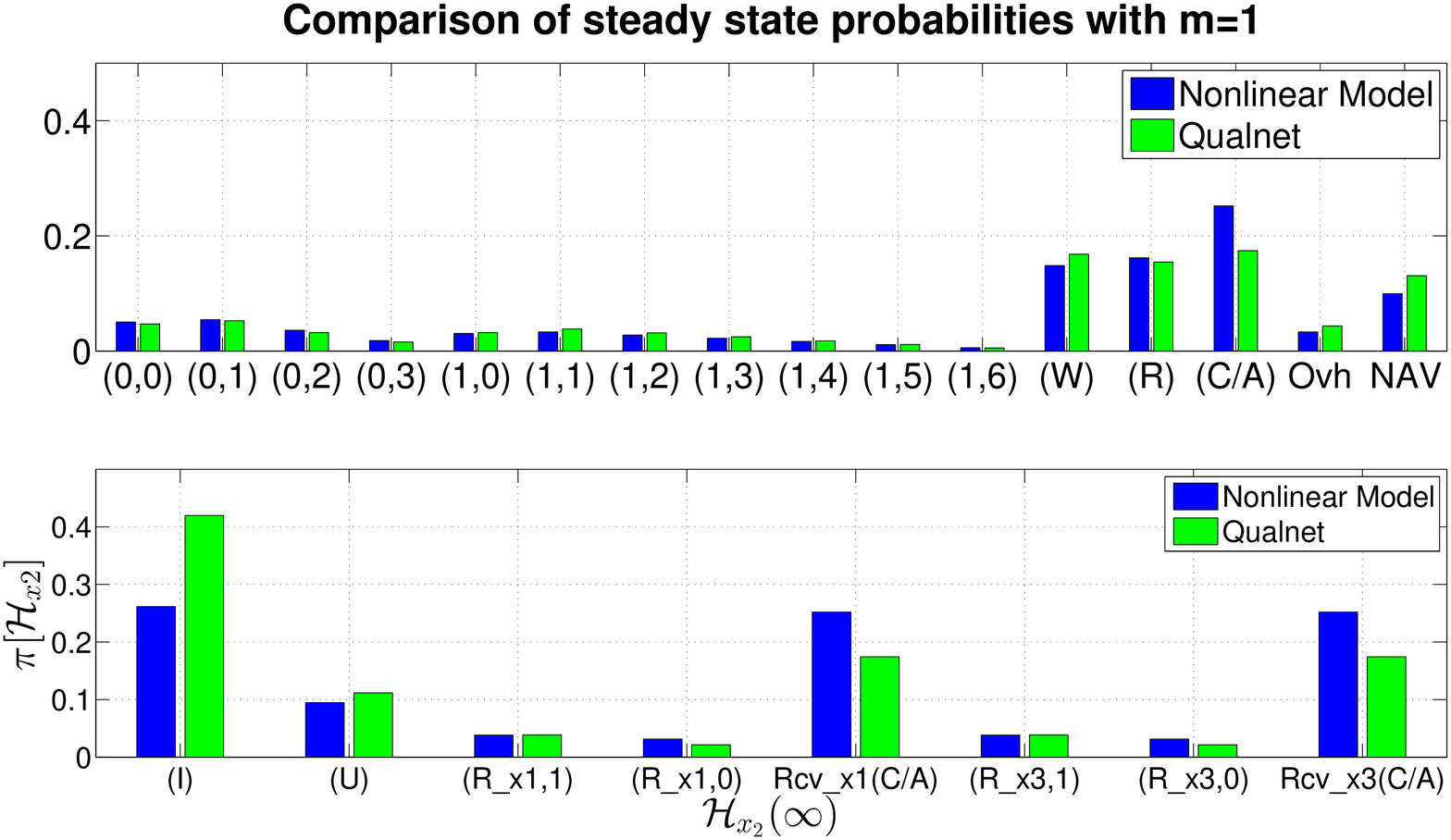} 
   \includegraphics[width=\textwidth]{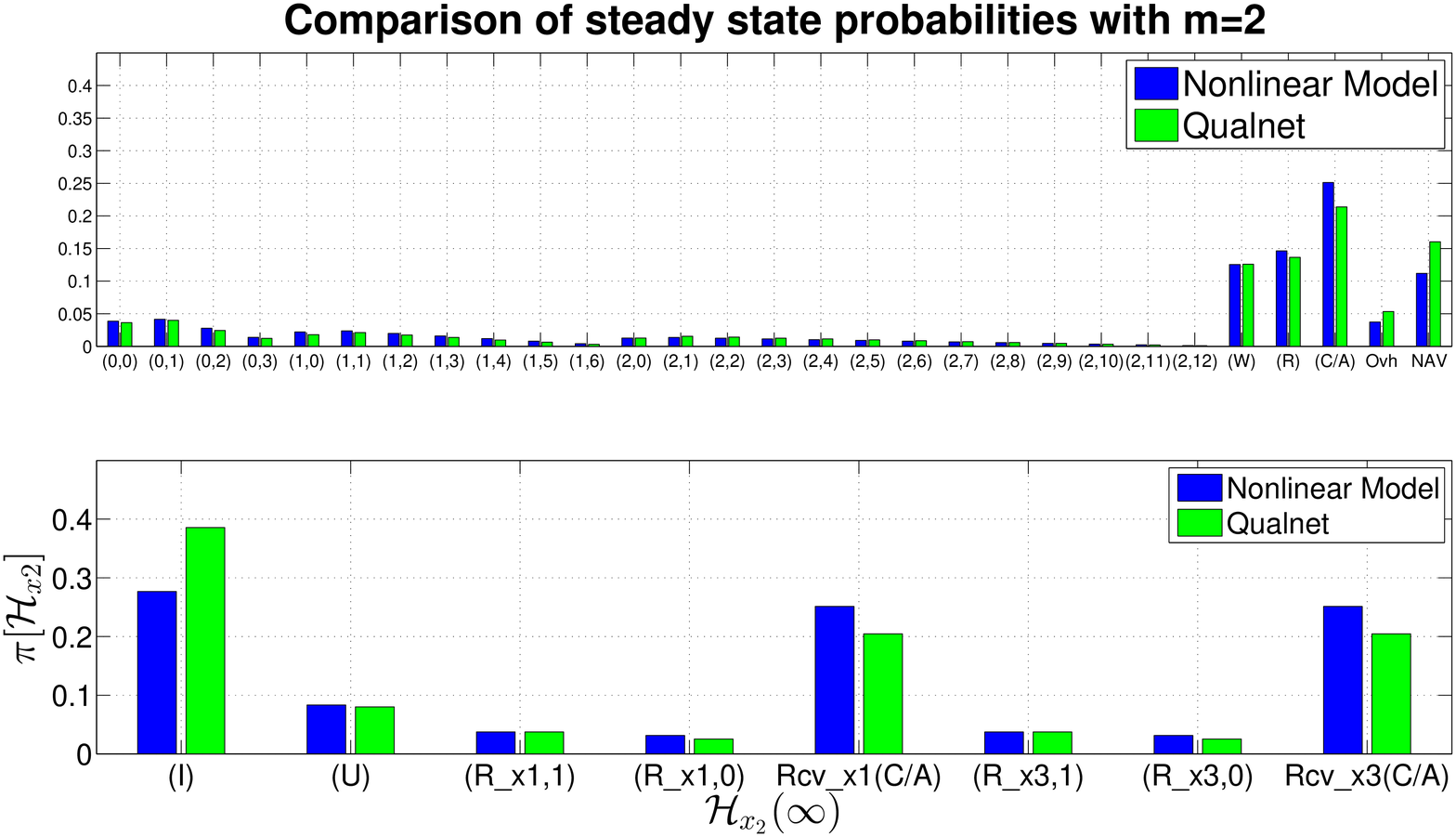} 
   \caption{\label{Ex3: 0-1-2 retry}
Comparison at $x_1$: the tuples denote (back-off stage, back-off
counter); 'Rcv' combines states of receiving RTS/ sending CTS/ receiving DATA;
'Ovh' denotes overhearing CTS. Comparison at $x_2$: the tuples represent the 1st
step or last step of receiving RTS from $x_1$ or $x_3$}
\end{figure}

The plots in Figure \ref{Ex3: 0-1-2 retry} show the difference between
analytical results and QualNet simulations. When RTS retransmission limit ($m$)
is $0$, it is possible to construct the stochastic joint state model of $x_1$,
$x_2$ and $x_3$. The corresponding results are shown by the red bar. The joint
model accurately predicts the behaviors of DCF. For the non-linear system, we
observe some significant deviation mainly due to the product approximation
approach as we bring closure to the system. However, as the complexity of the
system increases, i.e. $m=1,2$, the results improve. 

\section{Conclusion}
\label{sec: conclusion}
In this paper, we have introduced a new Markov model for the IEEE 802.11 Distributed
Coordination Function (DCF), a central mechanism of our wireless infrastructure. 
Our Markov model does not rely upon the assumption that collision probabilities
on each node are constant or independent of network topology.  Instead, we
have developed a detailed model of interconnected node states including
multiple back-off stages and binary exponential back-off counters to capture
the dominant first order effects of nodes' responses to contention.  The model is
complex, but it is necessarily so, and it is not so elaborate that it cannot be
analyzed.  Using the model, we have calculated
steady-state node states for two and three node networks including a
configuration that includes a hidden terminals with varying numbers of back-off
stages.  To determine the transition probabilities for steady-state
calculations we approximate the joint probability densities with marginal
probability densities using a product approximation. While this only uses a
small subset of the information available in the network description, we find
it sufficient to achieve excellent agreement with 
realistic simulations of network traffic.

\bibliography{DCF}

\begin{thebibliography}{10}

\bibitem{Qual2010}
{\em {QualNet 5.0.2 User's Guide}}, Scalable Network Technologies, Inc.,
  (2010), p.~388.

\bibitem{bianchi2000a}
{\sc Giuseppe Bianchi}, {\em {Performance analysis of the IEEE 802.11
  distributed coordination function}}, Selected Areas in Communications, IEEE
  Journal on, 18 (2000), pp.~535--547.

\bibitem{Dai2013}
{\sc Lin Dai and Xinghua Sun}, {\em {A unified analysis of IEEE 802.11 DCF
  networks: Stability, throughput, and delay}}, Mobile Computing, IEEE
  Transactions on, 12 (2013), pp.~1558--1572.

\bibitem{Daneshgaran2008}
{\sc F.~Daneshgaran, M.~Laddomada, F.~Mesiti, and M.~Mondin}, {\em {Unsaturated
  Throughput Analysis of IEEE 802.11 in Presence of Non Ideal Transmission
  Channel and Capture Effects}}, IEEE Transactions on Wireless Communications,
  7 (2008), pp.~1276--1286.

\bibitem{Foh2005}
{\sc CH~Foh and JW~Tantra}, {\em {Comments on IEEE 802. 11 saturation
  throughput analysis with freezing of backoff counters}}, IEEE Communications
  Letters, 9 (2005), pp.~130--132.

\bibitem{Garetto2008a}
{\sc Michele Garetto, Theodoros Salonidis, and Edward~W. Knightly}, {\em
  {Modeling per-flow throughput and capturing starvation in CSMA multi-hop
  wireless networks}}, IEEE/ACM Transactions on Networking, 16 (2008),
  pp.~864--877.

\bibitem{Garetto2005}
{\sc Michele Garetto, Jingpu Shi, and Edward~W. Knightly}, {\em {Modeling media
  access in embedded two-flow topologies of multi-hop wireless networks}}, in
  Proceedings of the 11th annual international conference on Mobile computing
  and networking - MobiCom '05, New York, New York, USA, Aug. 2005, ACM Press,
  p.~200.

\bibitem{Guillemin2011a}
{\sc Fabrice Guillemin, Charles Knessl, and Johan S~H van Leeuwaarden}, {\em
  {Wireless Multihop Networks with Stealing: Large Buffer Asymptotics via the
  Ray Method}}, SIAM Journal on Applied Mathematics, 71 (2011), pp.~1220--1240.

\bibitem{Hadzi-Velkov2003}
{\sc Z.~Hadzi-Velkov and B.~Spasenovski}, {\em {Saturation throughput - delay
  analysis of IEEE 802.11 DCF in fading channel}}, in IEEE International
  Conference on Communications, 2003. ICC '03., vol.~1, IEEE, 2003,
  pp.~121--126.

\bibitem{IEEE2012}
{\sc IEEE}, {\em {IEEE Standard for Local and metropolitan area networks, Part
  11: Wireless LAN Medium Access Control (MAC) and Physical Layer (PHY)
  Specifications}}, IEEE Std 802.11-2012,  (2012).

\bibitem{JaehyukChoi2006}
{\sc {Jaehyuk Choi}, {Joon Yoo}, and {Chong-Kwon Kim}}, {\em {A novel
  performance analysis model for an IEEE 802.11 wireless LAN}}, IEEE
  Communications Letters, 10 (2006), pp.~335--337.

\bibitem{Jang2012}
{\sc Beakcheol Jang and Mihail~L. Sichitiu}, {\em {IEEE 802.11 saturation
  throughput analysis in the presence of hidden terminals}}, IEEE/ACM
  Transactions on Networking, 20 (2012), pp.~557--570.

\bibitem{Mustapha2011}
{\sc I~Mustapha, JD~Jiya, and BU~Musa}, {\em {Modeling and Analysis of
  Collision Avoidance MAC Protocol in Multi-Hop Wireless Ad-Hoc Network}},
  \ldots Journal of Communication Networks and \ldots, 3 (2011), pp.~48--56.

\bibitem{Shi2012}
{\sc Zhefu Shi, Cory Beard, and Ken Mitchell}, {\em {Analytical models for
  understanding space, backoff, and flow correlation in CSMA wireless
  networks}}, Wireless Networks, 19 (2012), pp.~393--409.

\bibitem{Tinnirello2010}
{\sc Ilenia Tinnirello, Giuseppe Bianchi, and Yang Xiao}, {\em {Refinements on
  IEEE 802.11 distributed coordination function modeling approaches}},
  Vehicular Technology, IEEE \ldots, 59 (2010), pp.~1055--1067.

\bibitem{Tsertou2008}
{\sc a.~Tsertou and D.I. Laurenson}, {\em {Revisiting the Hidden Terminal
  Problem in a CSMA/CA Wireless Network}}, IEEE Transactions on Mobile
  Computing, 7 (2008), pp.~817--831.

\bibitem{Wu2002}
{\sc Haitao Wu, Yong Peng, and Keping Long}, {\em {Performance of reliable
  transport protocol over IEEE 802.11 wireless LAN: analysis and enhancement}},
  \ldots Conference of the IEEE \ldots, 00 (2002).

\bibitem{Wu2006}
{\sc Haitao Wu, Fan Zhu, Qian Zhang, and Zhisheng Niu}, {\em {WSN02-1: Analysis
  of IEEE 802.11 DCF with Hidden Terminals}}, \ldots , 2006. GLOBECOM'06. IEEE,
   (2006), pp.~2--6.

\bibitem{Zhai2003}
{\sc Hongqiang Zhai and Yuguang Fang}, {\em {Performance of wireless LANs based
  on IEEE 802.11 MAC protocols}}, \ldots , 2003. PIMRC 2003. 14th IEEE
  Proceedings \ldots,  (2003).

\bibitem{Ziouva2002}
{\sc Eustathia Ziouva and Theodore Antonakopoulos}, {\em {CSMA/CA performance
  under high traffic conditions: throughput and delay analysis}}, Computer
  communications, 25 (2002), pp.~313--321.

\end{thebibliography}
\bibliographystyle{siam}

\end{document}